\documentclass[3p]{elsarticle}

\usepackage{lineno,hyperref}
\modulolinenumbers[5]

\usepackage[algo2e,vlined,ruled]{algorithm2e}
\usepackage{algorithmicx}
\usepackage{adjustbox}
\usepackage{amsmath}
\usepackage{multirow}
\usepackage{subfigure}
\usepackage{caption}
\usepackage{fixmath}
\usepackage{makecell}

\usepackage{xcolor}

\usepackage{booktabs}

\newcommand{\sdc}{{\sf SDC}\xspace}
\newcommand{\lru}{{\sf LRU}\xspace}
\newcommand{\stdc}{{\sf STD}\xspace}
\newcommand{\stdclruf}{{\sf STD${\sf_{LRU}^{f}}$}\xspace}
\newcommand{\stdclruv}{{\sf STD${\sf_{LRU}^{v}}$}\xspace}
\newcommand{\stdcsdcv}{{\sf STD${\sf_{SDC}^{v}}$}\xspace}
\newcommand{\tstdc}{{\sf T${\sf_{SDC}^{v}}$}\xspace}

\journal{Information Processing \& Management}

\usepackage{numcompress}

\begin{document}

\begin{frontmatter}

\title{Topical Result Caching in Web Search Engines}


\author[ISTI]{Ida Mele} 
\ead{ida.mele@isti.cnr.it}

\author[ISTI]{Nicola Tonellotto}
\ead{nicola.tonellotto@isti.cnr.it}

\author[GT]{Ophir Frieder}
\ead{ophir@ir.cs.georgetown.edu}

\author[ISTI]{Raffaele Perego}
\ead{raffaele.perego@isti.cnr.it}

\address[ISTI]{ISTI-CNR, Pisa, Italy}
\address[GT]{Georgetown University, Washington, DC, USA}


\begin{abstract}
Caching search results is employed in information retrieval systems to expedite query processing and reduce back-end server workload. Motivated by the observation that queries belonging to different topics have different temporal-locality patterns, we investigate a novel caching model called \stdc (Static-Topic-Dynamic cache). It improves traditional \sdc (Static-Dynamic Cache) that stores in a static cache the results of popular queries and manages the dynamic cache with a replacement policy for intercepting the temporal variations in the query stream.

Our proposed caching scheme includes another layer for topic-based caching, where the entries are allocated to different topics (e.g., weather, education). The results of queries characterized by a topic are kept in the fraction of the cache dedicated to it. This permits to adapt the cache-space utilization to the temporal locality of the various topics and reduces cache misses due to those queries that are neither sufficiently popular to be in the static portion nor requested within short-time intervals to be in the dynamic portion.

We simulate different configurations for \stdc using two real-world query streams. Experiments demonstrate that our approach outperforms \sdc with an increase up to 3\% in terms of hit rates, and up to 36\% of gap reduction w.r.t. \sdc from the theoretical optimal caching algorithm.
\end{abstract}

%


\end{frontmatter}


\section{Introduction} \label{sec:introduction}

Caching is a fundamental architectural optimization strategy~\cite{barla:2015}, and query-result caching is critical for Web search efficiency. Query-result caching, as its name implies, stores the results of some selected user queries in a fast-access memory (cache) for future reuse. When a query is requested and we have a hit in the cache,  the cached results are directly returned to the user without reprocessing the request. Result caching improves the main efficiency-performance metrics of search engines, namely, latency and throughput. Indeed, serving the cached results decreases the latency perceived by the user issuing the query as well as avoids the usage of computational resources with a consequent improvement of the search engine throughput.

Another advantage of caching query results is the reduction of  energy consumption as  cached queries do not need to be reprocessed by the back-end servers.  Although other energy-efficiency optimization schemes exist~\cite{Catena:2015:LCP:2766462.2767809, tkde:catena, Pegasus, Teymorian:2013:RSQ:2505515.2505710}, these approaches are complementary to result caching and not a contradictory alternative.  Given desired and imposed ``Green Policy'' restrictions and the significant economic benefits due to the energy conservation, the interest of the search industry in energy saving is high.  Energy wise, the cost of a cached query is typically assumed to be nil, while a search of a query costs proportionally to its processing time in combination with the electricity price at the time of processing~\cite{Sazoglu:2013:FCM:2484028.2484182}.

The main challenge in query-result caching is the identification of those queries whose results should be cached. However, via query log mining, researchers observed high temporal locality in the query stream, enabling accurate search-engine side caching of \emph{popular} query results, i.e.,~results of queries frequently requested by different users~\cite{Saraiva:2001, Xie:2002}.  

The query-result cache can be \textit{static} or \textit{dynamic}. A static cache is periodically populated in an offline manner, with the results of past, most-popular queries. Query popularity is estimated observing previously submitted queries (e.g.,~previous day or week) in Web search logs; a simplifying, but not always correct, core assumption is that queries popular in the past remain popular in the future. A dynamic cache, as the name suggests, is dynamically updated; when the cache is full, an eviction/replacement policy is applied to decide which cache element must be removed to make space for the new one. The most common replacement policy for dynamic caches is the Least Recently Used (\lru) strategy:
every time a query is submitted, the cache is updated, keeping track of what query was used and when; if necessary, the cache entry used least recently is evicted to vacate space for the new entry.
The \lru strategy is effective without global knowledge and captures the ``bursty'' behavior of the queries by keeping recent queries in the cache and replacing those queries that are not requested for a long period of time. 

Static and dynamic caches can be combined together. Fagni et al.~\cite{Fagni:2006} proposed a Static-Dynamic Cache (\sdc) where the cache space is divided into two portions. The static portion stores results of the most popular queries, such as ``microsoft,'' ``youtube,'' or ``facebook.''  The dynamic portion maintains currency by applying the \lru strategy. This hybrid approach has proved successful in improving the performance of query-result caching with respect to both static and dynamic caching solutions in isolation.
Despite its good performance, \sdc suffers from some issues.  Static caching captures highly frequent queries, while \lru caching captures bursts of recently submitted queries. That is, static caching captures past queries that are popular over a relatively large time span (e.g.,~days or weeks) while \lru caching might fail to capture such long-term temporal locality, but does capture short-term popularity. However, a query might not be sufficiently globally popular to be cached in the static cache and not be requested so frequently to be kept in the dynamic cache, but it might become relatively popular in a specific time interval. For example, a query on a specific topic, such as weather forecast, is typically submitted in the early morning hours or at the end of a work day, but relatively seldom in the remaining hours of a day. 

We design a cache for query results that can adapt the cache-space utilization to the popularity of the various topics represented in the query stream. The intuition behind our approach is that queries can be grouped based on broad topics (e.g.,~the queries ``forecast'' and ``storm'' belong to the topic \texttt{weather}, while queries ``faculty'' and ``graduate'' to the topic \texttt{education}), and queries belonging to different topics might have different temporal-locality patterns. We assume that the \textit{topic popularity} is represented by the number of distinct queries belonging to the topic; to capture the specific locality patterns of each topic we  split the cache entries among the different topics proportional to their popularity. This provides queries belonging to frequently requested topics greater retention probability in the \lru cache. 

As an illustrative example, consider a cache with size 2 and the query stream  \texttt{abcadeafg}, where  query \texttt{a} is about a specific topic. A classical \lru strategy will get a 0\% hit rate (all queries will cause a miss). Instead, using 1 entry for the topic cache and 1 entry for the \lru cache we will get a 22.2\% hit rate (the first occurrence of \texttt{a} causes a miss, the other two occurrences will cause two hits in the topic cache).

\begin{figure*}[ht]
    \centering
    \includegraphics[width=.5\textwidth]{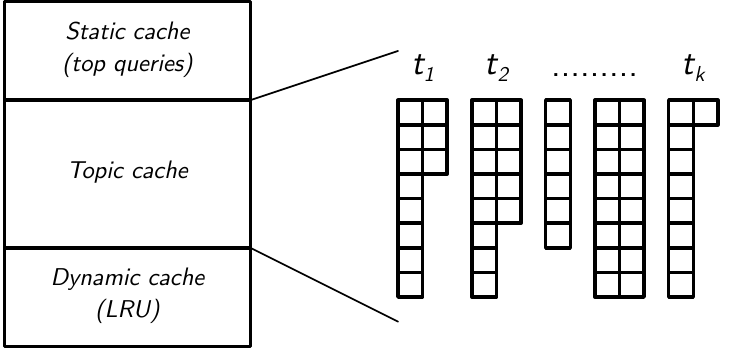}
    \caption{Example of \stdc cache.}
    \label{fig:STD_cache_example}
\end{figure*}

The topic cache can be combined with a static cache in different configurations. We propose to improve the \sdc approach by adding yet an additional cache space partition that stores results of queries based on their topics. We call our approach Static-Topic-Dynamic cache (\stdc) and in Fig~\ref{fig:STD_cache_example} is shown an example of it. To detect the query topics and incorporate them in the caching strategy we rely on the standard topic modeling approach called Latent Dirichlet Allocation (LDA)~\cite{Blei:2003:LDA}. LDA gets as input a document collection and returns lists of keywords representing the topics discussed in the collection. Each document in our setting consists of the query keywords and the textual content of their clicked results. Given the topics, the queries can be classified into topical categories, and  we estimate the topic popularity by observing the number of distinct queries belonging to that topic.

For our experiments, we use real-world query logs from AOL and Microsoft search engines, and we observe that modifying the caching strategy to include the topic partition increases the cache hit rate by greater than 3 percent with a consequent improvement of the query processing performance. This increase greatly reduces the gap w.r.t.   B\'el\'ady's optimal caching policy~\cite{Belady:1969}, measured as the difference between the hit rate of  B\'el\'ady's policy and the hit rate of \sdc/\stdc policies. While the \sdc gap is  ${\sim}10\%$, the \stdc gap is ${\sim}6\%$, with a relative improvement of more than $35\%$ when no cache admission policy is applied. This result is very good for query-result caching, considering that \sdc is the state-of-the-art and other attempts to enhance it resulted in small improvements~\cite{Kucukyilmaz:IPM2017}.

Compared to a traditional \sdc cache, we observe that our caching technique captures moderately popular queries not captured by the static cache; additionally we capture requests repeated within a large time interval that are likewise not captured by the dynamic cache. Some examples of queries that cause a miss in \sdc but not in \stdc are \textit{first bank, texas state bank, first national, renasant bank} for the topic \texttt{banking}, \textit{education world, everyday math, french-english dictionary} for the topic \texttt{education} and \textit{florida department of health, arthritis foundation, st vincents hospital, st marys hospital} for the topic \texttt{health}, just to mention a few of them.

The paper is structured as follows: Section~\ref{sec:rw} provides background information and discusses the related work in Web search engine caching, while Section~\ref{sec:methodology} introduces our \stdc cache model and different configurations of the topic-based cache, followed by how the topics are distilled. Then, Section~\ref{sec:datasets} describes and analyzes the query logs and document collections used for the assessment of the performance of our \stdc cache(s). Section~\ref{sec:experiments} details the experimental settings and reports on the results of our comprehensive evaluation. Finally, we conclude our investigation in Section~\ref{sec:conclusions}.

\section{Related Work}\label{sec:rw}

As background, we overview related efforts investigating query log mining, caching solutions in Web search engines, and query topic distillation.

\vspace{0.1cm}
\paragraph{Query Log Mining} Information extracted from query logs (e.g., search keywords, users' IDs, query timestamps, and clicked results) allows us to understand the behavior of users interacting with the search engine and, consequently, to enhance the effectiveness and efficiency of the retrieval systems. This desire for user behavioral understanding motivated researchers to investigate the mining of Web search logs in many different directions~\cite{silvestri09}. 

Raghavan and Sever~\cite{Raghavan:1995} performed a first study on how to use past knowledge to improve the effectiveness of information retrieval. They proposed to improve the result retrieval for queries similar to a \textit{query base} which is obtained from past optimal queries mined from query logs. 
Another approach exploits information extracted from query logs (e.g.,~query frequencies and distribution) for enhancing the performance of search engines, such as designing new caching strategies~\cite{Fagni:2006, Lempel:2003:PDC} and assessing the performance of caching techniques~\cite{Markatos:2001}. 
Other researchers analyzed query logs observing high correlations among query terms~\cite{Silverstein:1999} and common searching behavior among Web users (e.g.,~short queries)~\cite{Spink:IPM2006, Spink:2001}. 
A further analysis focusing on the temporal locality of queries was presented by Beitzel et al.~\cite{Beitzel:2004}. The authors analyzed one week of data, grouping queries per hours and per topic to see how their popularity changes during the day. They  observed that the query repetition rate is constant during the day, even though each popular query does not appear frequently in every hour. This result confirms the intuition that queries grouped by topic show some specific temporal localities in the stream.

\vspace{0.1cm}
\paragraph{Caching in Web Search Engines} 

Modern search engines are large-scale distributed systems where the inverted index is partitioned and stored among multiple back-end machines, each one running a search node. In addition, there is a front-end machine hosting the \textit{broker} for scheduling the queries among the different search nodes~\cite{Barroso:IEEEMicro2003}. 
When the query is submitted, the front-end machine sends its keywords to the different back-end machines, each one responsible of searching a given portion of the index. Then, it collects the results to recreate the Search Engine Result Page (SERP).

In modern search engines, query processing represents one of the major performance bottlenecks, so caching can help to speed up the search engine performance as well as to reduce the latency perceived by the users. Caching can be applied at different granularity including query results~\cite{Lempel:2003:PDC}, posting lists of query terms~\cite{Saraiva:2001}, and posting list intersections~\cite{Long:2005}. 
Saraiva et al.~\cite{Saraiva:2001} proposed a two-level architecture where the front-end machine caches the results of popular queries, while the back-end machines have a cache for the posting lists of most frequently requested terms. This architecture gets the benefits from both types of caching. Indeed, caching query results is faster since future requests of a query whose results are cached can be served immediately, without further processing, but it has a lower hit rate since a hit in cache occurs only with exact-matching queries. On the other hand, caching posting lists entails a better hit rate since the term overlap is greater, but the query needs to be processed by the back-end servers and the only saving is in the number of I/O operations over the disk storing the inverted index. 
Two-level caching was also studied by Baeza-Yates et al.~\cite{Baeza-Yates:SPIRE2003} in the static setting and by Altingovde et al.~\cite{Altingovde:2011} for dynamic setting. 

Three-level architectures store in the additional level of the cache the precomputed intersections of posting lists~\cite{Long:2005, Tolosa:2017,Zhou:2015}; this improves the processing of more complex queries such as AND or phrase queries. Ozcan et al.~\cite{Ozcan:2012fk} presented an even more sophisticated cache consisting of five levels for storing query results, precomputed scores, posting lists, precomputed intersections of posting lists, and documents. The caching approach is based on a greedy heuristic which allows to select the best item to cache. Each level of the cache has a priority queue, and the items are ordered in the queue according to their gains reflecting the frequency of the item's past accesses as well as its processing cost and storage overhead. 
We focus on a cache employed at the level of the front-end machine to store the results of most frequently requested queries.  
Our cache for query results can be used in combination with these multi-level cache architectures, as well. Improving the hit rate of the query-result cache is in fact beneficial for the lower layers of the caching system that provide an efficient processing of queries 
resulting in misses in the top-level cache. 

The query-result cache enhances the responsiveness of the search engine and saves resources. Depending on the space available, the cache can keep the entire HTML page (\textit{SERP cache})~\cite{Fagni:2006, Lempel:2003:PDC, Markatos:2001} or the URLs and/or snippets used for reconstructing it (\textit{DocID cache})~\cite{Saraiva:2001}. Caching the whole page of results requires more space and the same page cannot be used for different but similar queries; however, it is faster because when the query is requested, its cached SERP can be returned promptly. On the other hand, the \textit{DocID cache} allows to use the cache space in a more efficient way, but it is slower because even when we have a cache hit, some time would be needed to reconstruct the SERP. In~\cite{Altingovde:2011}, the authors proposed two levels of caching: first level for the SERPs and the second one for result documents of SERPs evicted from the first level cache. This allows to reconstruct the pages missing from the first level cache on the front-end server without processing the query on the back-end machines.  

The query-result cache can be managed in a \textit{static} or \textit{dynamic} way. A static cache is periodically (every day or week) populated with the results of frequent queries. Indeed, several authors observed that query frequency follows an inverse power law, which means that most of the queries are submitted a few times (e.g., they are long, with typos, or very rare), but a small portion of queries are requested several times and shared by different users~\cite{Fagni:2006, Lempel:2003:PDC, Xie:2002}. This justifies the use of a server-side static cache storing results of these queries. The cache is updated with the results of most popular queries observed in the previous period and for re-freshing the results. A dynamic cache changes the content of entries according to the query stream,  and it applies a replacement policy for deciding which element must be evicted when the cache is full. This policy tries to minimize the number of misses, and a popular one is Least Recently Used (\lru) which evicts the item not requested for the longer period of time.
Markatos~\cite{Markatos:2001} compared the hit rates achieved by \lru and its variations (e.g., Segmented LRU and LFU) over a query stream from the Excite query log. He observed that the \lru policy (and its variations) is very effective in query-result caching as it captures the temporal changes in the stream of requests. Although his work did not propose any novel caching algorithm, it showed the potentiality of caching in Web search.
Static and dynamic caches can be combined to improve the hit rate. Fagni et al. in~\cite{Fagni:2006} presented Static-Dynamic Cache (\sdc), consisting of a static portion for storing the results of most frequent queries and a dynamic portion for which an \lru-like approach is applied. Combined or not with prefetching strategies, \sdc outperforms purely static or dynamic caching policies. We present a further improvement of \sdc by adding another layer which is made of several topic-dependent \lru caches.

To improve cache performance, the system can apply an admission policy which accepts to the cache only those queries with a high re-submission probability. The idea behind these policies is that caching search results of queries that are not requested anymore or are requested after a long time (and generally after their eviction from the cache) should be avoided. Indeed, these queries (also called ``polluting'' queries~\cite{baeza:SPIRE2007admission}) would only waste space in the cache, so it would be better to just ignore them and give more chance to the results of other queries to be kept in the dynamic portion of the cache (e.g., the \lru queue). 
These policies mostly rely on features extracted from the queries or from usage information~\cite{baeza:SPIRE2007admission, Gan:2009, Kucukyilmaz:IPM2017}. Baeza-Yates et al. proposed an admission policy for restricting unfrequent and too long queries from entering the cache~\cite{baeza:SPIRE2007admission}. In~\cite{Kucukyilmaz:IPM2017}, Kucukyilmaz et al. proposed a machine learning approach for predicting the next  
queries to cache. Their approach is based on features extracted from the query, index, term frequency, query frequency, and user session. This policy on top of \sdc yields to a small improvement of the hit rate compared to the one achieved with our novel caching model. Recent works employed machine learning to predict queries not worthy to be cached~\cite{Ozcan:TWEB2013} or to adjust the time-to-live values of queries~\cite{Alici:2012}.
Alternatively, the policy can favor queries that are more costly to process~\cite{Ozcan:2009}.
Other works focus on predicting stale queries~\cite{Jonassen:2012} and use the information extracted from both queries and session features to increase the hit rate of the cache.
We do not investigate novel admission policies; rather we use them in coordination with our caching model as they allow to enhance the input stream of queries. We followed ~\cite{baeza:SPIRE2007admission} and used the \textit{stateful} and \textit{stateless} features of queries for implementing the admission policy. The former is given by statistical information computed over the past stream (e.g., the number of times the query was submitted in the past). The latter is based on the query characteristics (e.g., the length of the query). The idea is to avoid the caching of infrequently requested queries; long queries are typically rare and are unlikely to be resubmitted.
Moreover,  we computed an upper bound of the best hit ratio that can be achieved using an admission policy with a cache of a given size. For this reason we used the B\'el\'ady's optimal caching policy~\cite{Belady:1969} that assumes to know the future and evicts the results of the query not requested for the longest time (representing the best replacement policy) and an oracle which knows the future and does not admit in the cache the singleton queries (representing the best  admission policy).

Orthogonal research works focus on caching single results (e.g.,~URLs and snippets) that are used for reconstructing the final page of results of queries (SERP). Cambazoglu et al.~\cite{Cambazoglu:TWEB2012} present techniques for computing result pages of unseen queries starting from the cached results of previously requested query. Ceccarelli et al. investigate the problem of caching query-based snippets \cite{edbt2011}. In~\cite{Anagnostopoulos:TOIS2015}, the \textit{stochastic query covering} approach is presented for selecting the set of documents to store in the cache. These documents serve many queries and can be used for fast retrieval as well as for approximating the results of a query when the connection between the front-end server with the back-end machines is too slow or unavailable. We do not aim at improving document caching as our strategy for query-result caching is orthogonal to it and can be employed together with a document cache in a two-level fashion architecture as in~\cite{Altingovde:2011}. So, the first level caches the SERPs of queries, while the second level caches the  documents used for reconstructing the SERPs. When one of the SERP is evicted from the first level cache, the system may decide to cache its documents in the second level cache, so that queries evicted from the first level cache have a second chance to be served by cached documents stored in the second level cache.

Another line of research focuses on index tiering~\cite{Kayaaslan:IPM2013} as well as on document replication~\cite{Daoud:IPM2016}; these techniques can be employed in synergy with server-side caching with the purpose of improving the performance of distributed search engines.

\paragraph{Query-Topic Distillation} Two challenging problems in query log mining are query classification and query-topic distillation. 
Flat and hierarchical taxonomies were proposed for classifying the user queries, even though these taxonomies are limited to some specific domains and provide a broad categorization~\cite{Broder:2007, Spink:2001}. 
Query-topic distillation/detection can improve quality of the Web search~\cite{Ophir05}. It allows to distill the multiple possible topics behind an ambiguous/broad query which is beneficial for better understanding the query intention and improving the automatic reformulation of queries~\cite{He:WWW2007}. Nevertheless, query-topic detection is not easy due to the shortness and lack of context of queries. One possible approach for understanding the topic of a query is enriching its search keywords with the content (from snippet or page) of its top results~\cite{He:WWW2007}. Another solution consists of using the content of only the clicked results. As  explained in Sec.~\ref{sec:methodology}, we opted for this second solution since the user click is a strong indicator of the relevance of the document to the query. Once the search queries are enriched with page content, the resulting document collection can be used to discover the topics. Text clustering techniques or topic modeling approaches can be applied. Topic models have been used for several years as they allow to discover the topics discussed in large document collections~\cite{Blei:2003:LDA, Li:IPM2018, Rashid:IPM2019}. We opt to use a topic-model approach (e.g.,~LDA~\cite{Blei:2003:LDA}) as it is completely unsupervised and domain independent, plus it leverages the word co-occurrences, providing better results compared to a basic text clustering approach~\cite{Lu:2011}.

\section{Methodology} \label{sec:methodology}

We propose a new query-result caching strategy based on user search topics. 
We first describe our caching architecture, including possible implementation configurations, and continue by discussing our query topic extraction approach. We assume that the cache stores the SERPs of the queries; onward we use \textit{query results} or just \textit{results} to refer to the content of a cache entry. 

\subsection{Topic-based Caching} 
Given the total number $N$ of cache entries available for storing the results of past queries, we propose a Static-Topic-Dynamic (\stdc) cache  which includes the following components: 
\begin{itemize}
    \item a \textit{static} cache  $\mathcal{S}$ of size $ |\mathcal{S}| = f_s \cdot N$ entries, used for caching the results of the most frequently requested queries. The static cache $\mathcal{S}$ is updated periodically  with the fresh results of the top frequent $|\mathcal{S}|$ queries submitted in the previous time frame (e.g.,~the previous week or  month). This static cache is expected to serve very popular queries such as navigational ones (e.g.,~``google'' and ``facebook'');
    \item a \textit{topic} cache  $\mathcal{T}$ of size $|\mathcal{T}| = f_t \cdot N$ entries, which is in turn  partitioned in $k$ topic-based sections $\mathcal{T}.\tau$, with $\tau \in \{t_1, t_2, \ldots, t_k\}$, where $k$ is the number of distinct topics. Each section $\mathcal{T}.\tau$ is considered as an independent cache, managed with some caching policy (e.g.,~\lru or \sdc), and  aimed at capturing the specific temporal locality of the queries belonging to a given topic, i.e.,~queries more frequent in specific time intervals or with periodic ``burstiness'' (e.g.,~queries on weather forecasting, typically issued in the morning, or queries on sport events, typically issued in the weekend);  
     \item a \textit{dynamic} cache $\mathcal{D}$ of size $|\mathcal{D}| = f_{d} \cdot N$. The dynamic cache $\mathcal{D}$ is managed using some replacement policy, such as \lru.  It is expected to store the results of ``bursty'' queries (i.e.,~queries requested  frequently for a short period of time) that are not captured by neither $\mathcal{S}$ nor $\mathcal{T}$ as they are not sufficiently popular or are unassigned to any of the $k$ topics. Queries cannot be assigned to a topic for two reasons: (i)~the query was never seen before, hence the topic classifier fails to detect its topics, or (ii)~even though it was already submitted in the past, no topic was assigned to it due to a very low classification confidence (see Sec.~\ref{sec:lda}). 
\end{itemize}

The parameters $f_s$, $f_t$, and $f_d$ denote the fractions of entries $N$ devoted to the static, topic, and dynamic caches, respectively, so that $f_s + f_t + f_d = 1$. Note that, if $f_t = 0$, our \stdc cache becomes the classical \sdc cache.
The number of entries in each section $\mathcal{T}.\tau$ of the topic cache can be fixed, i.e.,~$|\mathcal{T}.\tau| = |\mathcal{T}|/k$, for every $\tau \in \{t_1,\ldots, t_k\}$, or chosen on the basis of the popularity of the associated topic (observed in a past query stream). In the latter case, we model the topic popularity as the number of distinct queries in the topic, since estimating this number allows us to assign to the topic a number of entries proportional to its requested queries. This entails a more efficient utilization of the cache space since queries belonging to a popular topic have greater chances to be retained in the cache as their topic receives more entries as compared to other queries belonging to unpopular topics.

\begin{figure*}[ht]
    \centering
    \includegraphics[width=.99\textwidth]{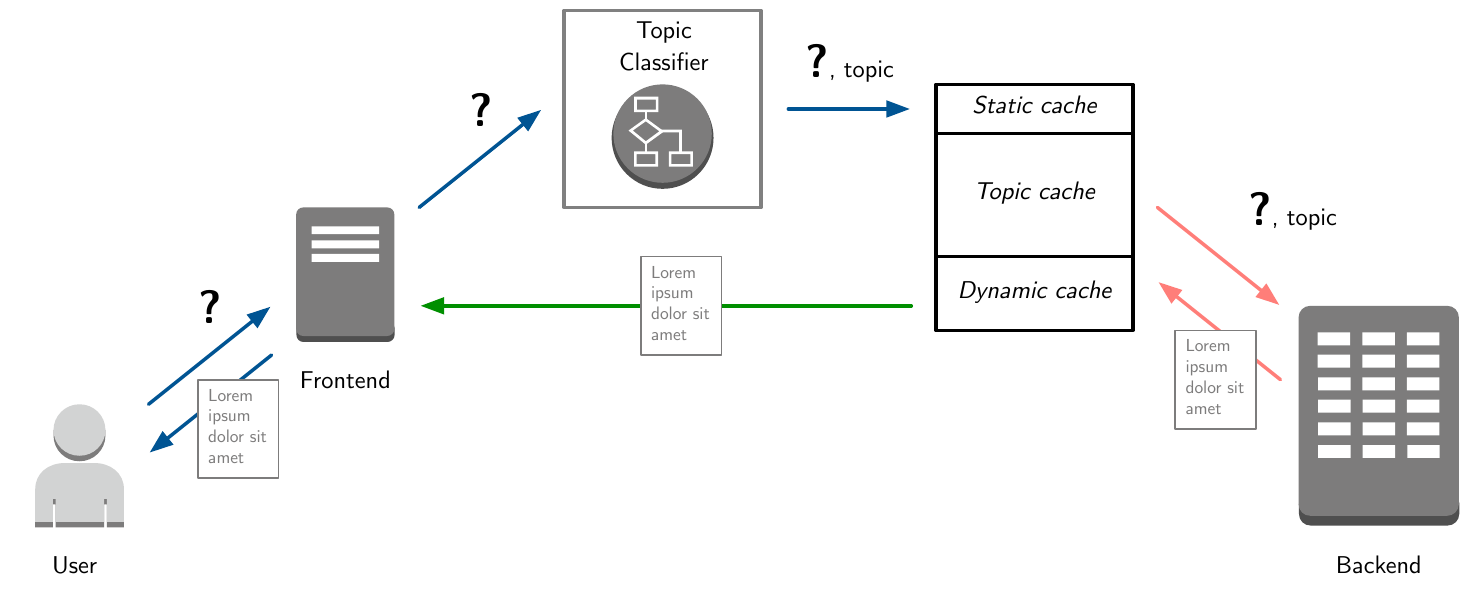}
    \caption{Example of a search engine using the \stdc cache. The question mark represents the user query. In case of a hit in the cache, the results are returned to the user immediately (green arrow). Otherwise, a cache miss is encountered and the request is sent to the back-end server (red arrows).}
    \label{fig:system_example}
\end{figure*}

In Fig.~\ref{fig:system_example} we show an example of how the proposed cache can be employed in a search engine system. The management of the \stdc cache is reported in Alg.~\ref{algo:std}. When a query with its topic $\tau \in \{t_1,\ldots,t_k\}$ arrives, the cache manager first checks if the query is in the static cache $\mathcal{S}$. If so, we have a hit; otherwise, if the query has a topic handled by the cache, the manager checks the topic-specific section of the topic cache $\mathcal{T}.\tau$, updating the topic cache with its specific replacement policy, if necessary, and producing a hit or a miss if the query was cached or not. If the query was not assigned to any topic, the dynamic cache $\mathcal{D}$ is responsible for managing the query and producing a hit or a miss. We note that the pseudo-code does not detail the retrieval of the query results from the cache or its processing on the inverted index of the search engine in case of hit or miss, respectively.

\begin{algorithm2e}[h!]
	\DontPrintSemicolon
	\SetKwInOut{Input}{Input}
	\SetKwInOut{Output}{Output}
	\Input{$~~~$ A query $q$ and its topic $\tau$}
	\Output{$~~~$ Hit or miss and updated \stdc cache}
	{
	    \nl \If{$q$ {\bf in } $\mathcal{S}$}{
	        \nl {\bf return } hit\;
	    } 
	    \nl \If{$\tau$ {\bf in } $\{t_1, \ldots, t_k\}$}{
	        \nl \If{$q$ {\bf in } $\mathcal{T}.\tau$} {
	            \nl update $\mathcal{T}.\tau$ and {\bf return } hit\;
	        } \Else {
	            \nl update $\mathcal{T}.\tau$ and {\bf return } miss\;
	        }
	    }
        \nl \If{$q$ {\bf in } $\mathcal{D}$} {
            \nl update $\mathcal{D}$ and {\bf return } hit\;
        } \Else {
            \nl update $\mathcal{D}$ and {\bf return } miss\;
        }
	}
	\caption{\textbf{The \stdc cache management process.}}\label{algo:std}
\end{algorithm2e}

\newpage
Note that the cache misses incur different costs since some queries are more expensive to process than others (in term of time and resources). For our research purpose of comparing different caching architectures, however, we simplify the performance analysis and focus on the hit rate, considering all the misses with the same cost. In other efforts that focus on determining which element must be evicted from the dynamic cache or admitted to the static cache, the cost of the misses is taken into account~\cite{Ozcan:2009, Gan:2009, Ozcan:TWEB2013}. Anyway, these strategies, based on how costly is the computation of query results, can be used in synergy with our caching architecture to improve the overall performance.

\subsection{\stdc Cache Configurations}
\begin{figure*}[ht]
    \centering
    \includegraphics[width=.95\textwidth]{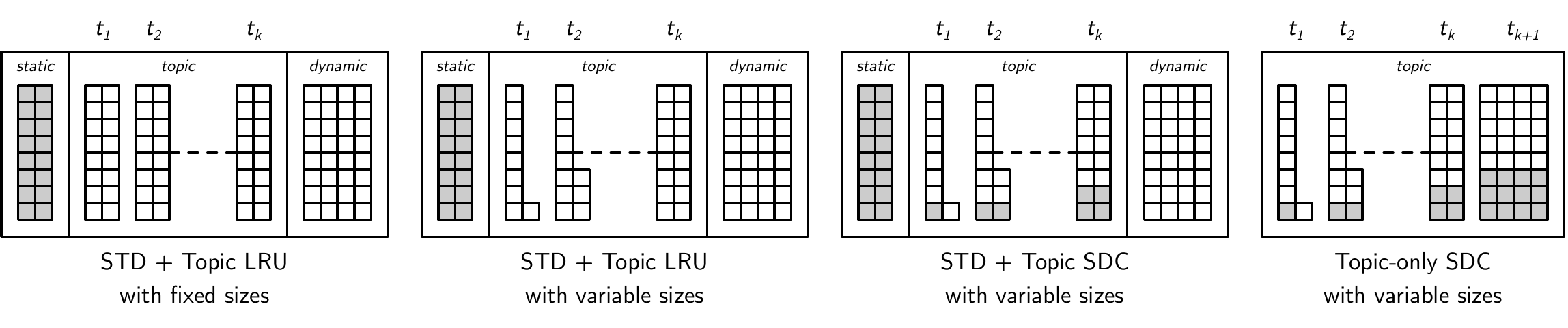}
    \caption{Different Configurations of a \stdc cache.}
    \label{fig:SDC_Topics_caches}
\end{figure*}

Our proposed \stdc cache model can be implemented in different ways, depending on several parameters such as the values of $f_s$, $f_t$ and $f_d$, the number of entries assigned to each topic in the topic cache, the replacement policy adopted, and so on. In the following, we illustrate some of these implementations (see Fig.~\ref{fig:SDC_Topics_caches}), which will be part of our experimental evaluation (see Sec.~\ref{sec:experiments}).

\begin{itemize}
\item \stdc with topic cache managed by \lru with fixed size (\stdclruf). This cache includes the static, topic, and dynamic caches discussed above. The topic cache entries are divided equally among the different topics (i.e.,~without taking into account the topic popularity), and each topic cache section is managed according to the \lru replacement policy.
 
\item \stdc with topic cache managed by \lru with variable  size (\stdclruv). This cache is similar to the previous one with the difference that each topic has a number of entries proportional to its popularity. We quantified this topic popularity as the number of distinct queries that belong to the topic in the training set of the query log.

\item \stdc with topic cache managed by \sdc with variable entry size (\stdcsdcv). This cache is similar to the \stdclruv cache, but now the topic cache is managed by \sdc instead of \lru. 
Each topic gets a given number of entries proportional to its topic popularity, and all topic cache sections are split in a static and dynamic cache. The fraction of entries allocated to the static portion of these caches is a constant fraction of the topic cache entries and denoted with $f_t^s$. The remaining entries, allocated for the topic, are managed by \lru. 

\item Topic-only cache managed by \sdc with variable entry size (\tstdc). This is an alternative version of the previous implementation since the queries with no topic are managed as queries belonging to an additional topic $k+1$. This means that instead of having a predefined size for static and dynamic caches, the number of entries would depend on the number of queries without a topic.
\end{itemize}

\subsection{Modeling Queries as Topics}\label{sec:lda}

Query topic categorization of user queries is well addressed within Web companies to increase effectiveness, efficiency, and revenue potential in general-purpose Web search engines~\cite{Ophir05}. Unfortunately, we cannot rely on the query topic characterization provided by the search engine industry since this information is missing from the query logs we use. Thus, to distill the topics of the queries in our query logs, we rely on LDA topic modeling. 

\paragraph{Latent Dirichlet Allocation (LDA)~\cite{Blei:2003:LDA}}  LDA is an unsupervised approach not requiring any prior knowledge of the domain for discovering the latent topics. Given a collection of documents and  the number $k$ of topics, it returns $k$ lists of keywords, each representing a latent topic.
Let $\theta_d$ be the per-document topic distribution, which is assumed to be drawn from a Dirichlet distribution with hyper-parameter $\alpha$. The documents are a mixture of topics, and the multinomial random variable $z_{d,n}$ of a topic to appear in position $n$ of document $d$ is conditioned on $\theta_d$. Words appearing in a document are selected according to another multinomial distribution with hyper-parameter $\beta$, conditioned on the chosen topic. In this way, each word has a probability that depends on its likelihood to appear in the document relevant to the topic. In summary, LDA can be seen as a generative process where documents are generated sequentially as reported in Alg.~\ref{algo:lda}.

\begin{algorithm2e}[h!]
	\DontPrintSemicolon
	{
        \nl \For{\emph{{\bf each} document} $d$} {
	        \nl Draw $\theta_d \sim \text{Dir} (\alpha)$\;
		    \nl \For{\emph{{\bf each} word position} $n$} {
                \nl {Draw} a topic $z_{d,n} \sim  \text{Multinomial} (\theta_d)$\;
			    \nl {Draw} a word $w_{d,n} \sim \text{Multinomial} (\beta | z_{d,n})$\;
	        }
	    }
	}
	\caption{\textbf{The LDA generative process.}}\label{algo:lda}
\end{algorithm2e}

By inverting the generative process, it is possible to infer the topics from the words appearing in the documents. So, given a document $d$, we have to compute the posterior distribution of the hidden variables $\mathbf{z}_{d}$ and $\theta_d$ as follows:
\begin{equation*}
p(\theta_d, \mathbf{z}_d | \mathbf{w}_d, \alpha,\beta) = \frac{p(\theta_d, \mathbf{z}_d, \mathbf{w}_d | \alpha, \beta)} {p(\mathbf{w}_d | \alpha, \beta)},
\end{equation*}
where the vector $\mathbf{w}_d$ represents the words observed in $d$, while the vector $\mathbf{z}_d$ represents the positions of words in $d$. Both vectors have the same size, equal to the length of $d$.
Statistical inference techniques, such as Gibbs sampling~\cite{Griffiths:2004}, are employed to learn the underlying topic distribution $\theta_d$ of each document.

\paragraph{Finding Latent Topics from Query-Document Pairs} Given a training query log, we aim at learning a query topic classifier based on LDA. Since queries are short and lack context, it is difficult to train the model accurately. To circumvent this problem, we enrich the queries with the content of their clicked pages whose URLs are available in the training query log. We thus create a collection of query-document samples made of queries plus the text of their clicked results gathered from the Web. In case, for a given query, the user did not click any results, or the clicked URL was not available any longer, we remove the corresponding query from the set. Given a query-document pair, we use this content as a proxy of the query, and train LDA to learn the topics discussed in the collection of query-document pairs.

The trained LDA classifier returns a distribution of topics for each query-document pair. Since we assume that a query can be assigned to only one topic, we always choose, for the pair, the topic with the highest probability. In our experiments, we use different query logs w.r.t. the training one, and we assume the LDA classifier is able to classify only queries already seen in the training query log, since for new queries we will lack the content of clicked pages.

\paragraph{Query Topic Assignment} Once we know the topics of the proxy query-documents, we must assign a single topic to each query. Since the same query might appear in different query-document pairs, possibly assigned to different topics, we must decide which one of these topics should be associated with the query. To this end, we adopted a simple voting scheme that assigns to each query the topic of the query-document that got more clicks by the users. In doing so, we leverage the strong signal coming from clicks about the relevance of a document and its topic to the information need expressed by the query. Also, it allows us to estimate the most popular topic that can be assigned to ambiguous queries (i.e.,~queries with more than one meaning that have more possible topics).

\paragraph{Estimating Topic Popularity} \label{sec:methodology_sizes}
In some of the implementations of our topic cache (i.e., \stdclruv, \stdcsdcv, \tstdc), we assign to each topic an amount of cache entries proportional to the topic popularity. Similarly to the static cache, where past popular queries are assumed popular in the future, we assume that popular topics observed in the past  remain popular in the future; so they get more entries in the topic cache $\mathcal{T}$. 

We quantified this topic popularity as the number of distinct queries that belong to the topic $q_{\tau}$. Note that this statistic is computed over the training period. More in detail, let $|\mathcal{T}|$ be the size of the topic cache, $q$ be the number of distinct queries in the training set, each topic $\tau$ will get a number of entries $|\mathcal{T}.\tau|$ equal to 
\begin{equation*}
    |\mathcal{T}.\tau| = \left\lfloor  \frac{|\mathcal{T}|}{q} \cdot q_{\tau} \right\rceil
\end{equation*}

\noindent where the symbol $\lfloor x \rceil$ is the nearest integer to $x$. For example, suppose we have a topic cache with size $|\mathcal{T}| = 5$ and 9 distinct queries observed in the training: 6 for the topic \texttt{weather} and 3 for \texttt{education}, we will have $|\mathcal{T}.{\sf weather}| = \lfloor 3.33 \rceil = 3$ and $|\mathcal{T}.{\sf education}| = \lfloor 1.66 \rceil = 2$.

\section{Dataset Description}\label{sec:datasets}

\paragraph{Query logs} For our experiments, we used two real-world query logs, namely, the AOL and MSN query logs.

The AOL query log~\cite{pct-ps-infoscale06} consists of about $36M$ of query records, submitted by $650K$ users over a period of three months (from March to May 2006). Each record in the dataset is made of the user ID, the query keyword(s), the timestamp, the rank and the URL of the domain of the clicked result. The last two fields represent a click-through event, and they are present only if the user clicked on a search result. About $19.4M$ records have a domain URL, and the number of distinct URLs is $1.6M$. 
Notice that, when a user clicked on more than one result in the same search session, there are repeated records for that query (one for each clicked result). Hence, for our experiments, we recreated the query stream by removing the  duplicated records  representing multiple clicks for the same query, and we kept only the ``first query'' of the sequence. The final query stream, used for the simulation of the cache, consists of $20M$ queries. 

The MSN query log~\cite{Craswell:MSN:2009} consists of about $14.9M$ records storing the timestamp, the search keyword(s), the query id, the session id, and the result count. 
About $8.8M$ records have a clicked URL for a total of $3.4M$ distinct URLs.

\begin{figure}
\centering
\subfigure[AOL\label{fig:distr_AOL}]{\includegraphics[scale=0.5]{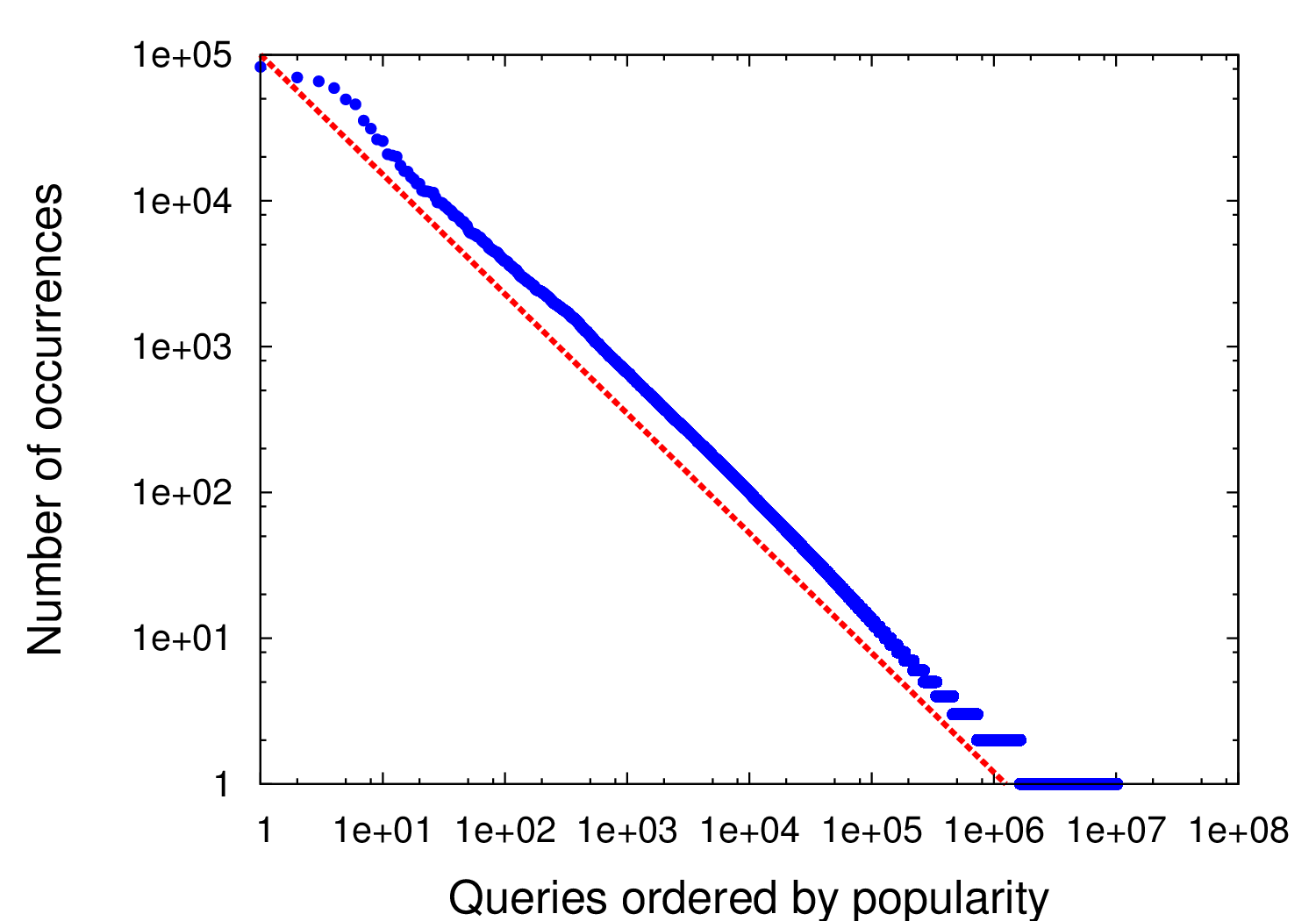}}
\subfigure[MSN\label{fig:distr_MSN}]{\includegraphics[scale=0.5]{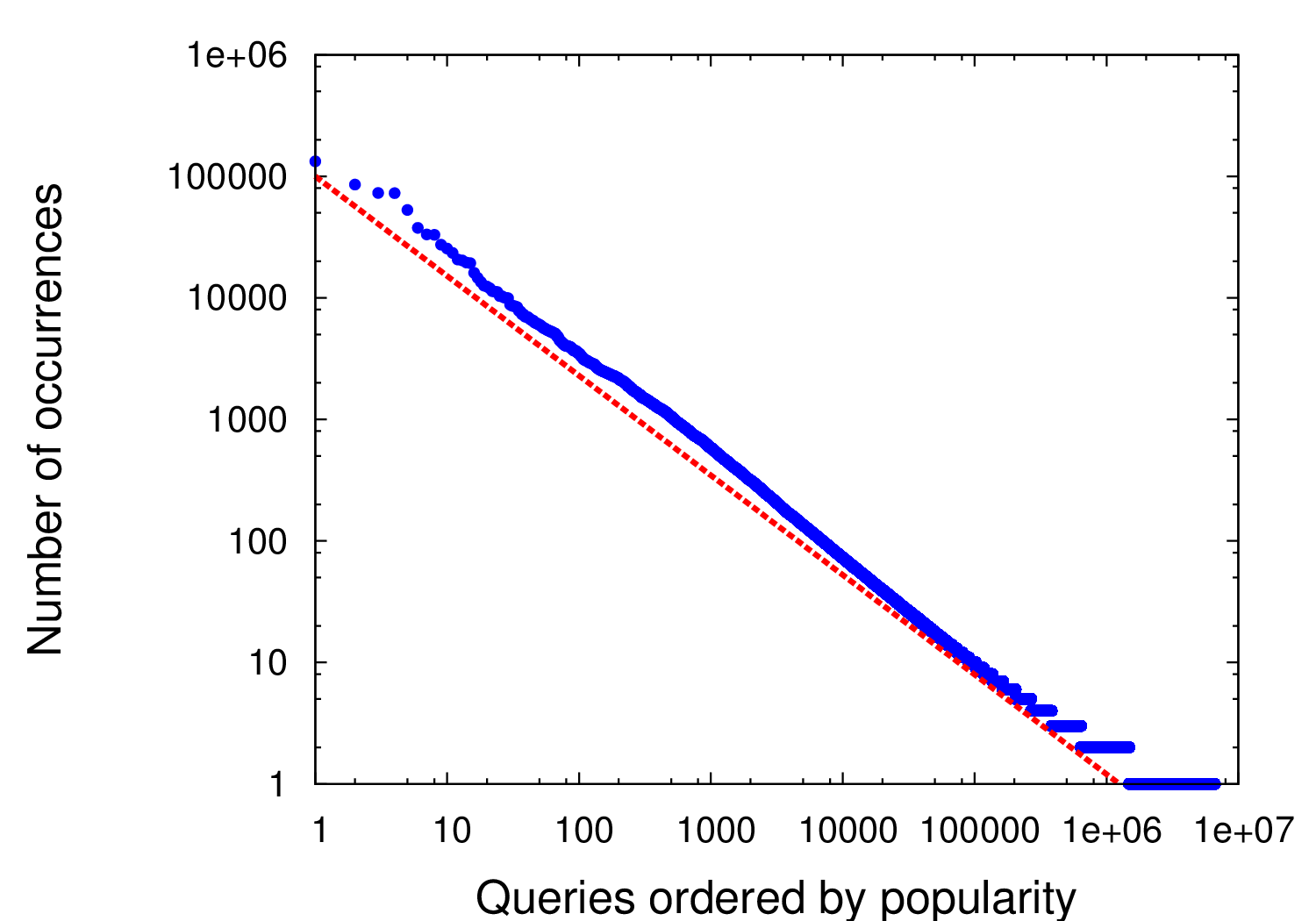}}
\caption{Query logs: Distribution of the query frequencies.}
\label{fig:distr}
\end{figure}

For both datasets, we preprocessed the queries by removing special characters and converting them to lowercase. After the preprocessing, the number of distinct queries is $9.3M$ and $6.2M$ for AOL and MSN, respectively. 
The distribution of the query popularity follows a power law in both query logs as shown in Fig.~\ref{fig:distr}. We plot the distribution using a log-log scale; the queries are ordered along the $x$-axis by popularity, while the number of occurrences of the query (i.e.,~query frequency) is shown on the $y$-axis.

\vspace{0.5cm}
We partitioned both query logs into two portions, one for training purposes (e.g.,~cache initialization) and one for testing the performance of the cache. We sorted the query logs by time and split them into two fractions: $X$ for the training set and $100 - X$ for the test set with $X=30\%, 50\%$, and $70\%$. 
We present the results for the $70\%$-$30\%$ split for caching without admission policies, but we observed similar results for the other training-test splits. In this case, the training (resp. test) set contains $6.7M$ (resp. $3.2M$) unique queries for AOL and $4.5M$ (resp. $2.1M$) unique queries for MSN.
We also tried different splitting of the datasets, such as weekly splits, but the average results were not much different from the ones reported in the next section.

\paragraph{Document Collections} Given the URLs from both query logs, we collected the associated pages from the Web, and gathered $1M$ documents for AOL and $2.1M$ documents for MSN. Then, we extracted and pre-processed the text (e.g.,~stop-word removal, lemmatization, and stemming). We removed overly short and long documents (less than 5 and more than 100K words). Lastly, we enriched the documents with the corresponding query keywords. 

\paragraph{LDA Topics} To train the LDA model we subsampled the documents from the training period and used $500K$ documents from AOL and $350K$ documents from MSN query log. Notice that this training is only used for learning the topics discussed in the collection (e.g., sport, politics, weather). After the topic detection, given a query and its clicked result we use LDA to predict the topic of the query.

We removed the very frequent and rare words from the dictionary, and we set the number $k$ of topics to discover to $500$, estimated empirically. The approach is probabilistic, hence the topic detection can change with different collections and different number of topics. We tried other configurations, changing the subsets of documents in the training set and using different values of $k$ (e.g., 50, 100, and 500); we observed that the impact on the caching performance was negligible. Some of the topic keywords, extracted from the AOL dataset, are shown in Table~\ref{tb:topics}.

\begin{figure}[bp!]
\centering
\includegraphics[width=.45\textwidth]{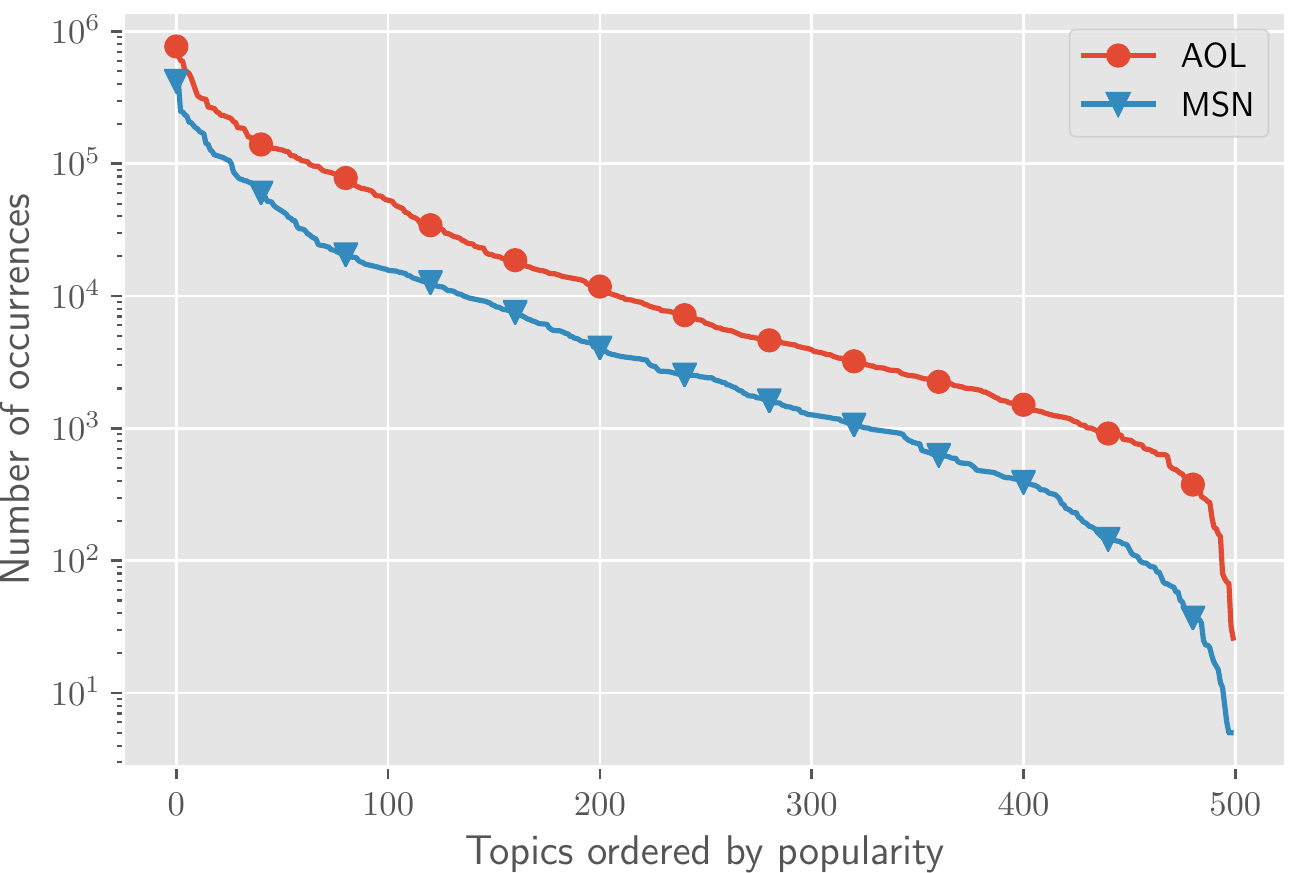}
\caption{Distribution of topic popularities for AOL and MSN query logs.}
\label{fig:topicDistr}
\end{figure}

In Fig.~\ref{fig:topicDistr}, we report the distribution of topics extracted from both datasets. It is worth noting that the topic portion of \stdc cache exploits the subset of queries in the test set stream having a known topic. These queries are necessarily among those already encountered in the training set stream and successfully classified. The test queries that were not assigned to a topic compete instead for the use of the static and dynamic portions of the \stdc cache. The percentage of queries in the test set with a topic is $65\%$ for AOL and $58\%$ for the MSN query log.

\begin{table}[t!]
\centering
\caption{Some topics and their keywords extracted from the AOL search queries and the page content of their clicked results.} 
\begin{tabular}{c|c} 
\toprule
\textbf{Topic} & \textbf{Topic Keywords}\\
\midrule
\textbf{shopping} & shop, order, item, ship, gift, custom, sale, return, account, cart\\
\textbf{university} & student, program, faculti, campu, graduat, research, academ, alumni, colleg, univers\\
\textbf{weather} & weather, forecast, snow, storm, rain, wind, winter, radar, flood, cold \\
\textbf{movies} & movi, comic, news, star, theater, review, marvel, film, seri, comedi\\
\textbf{cooking} & recip, cook, bean, chicken, chef, salad, cake, flavor, potato, rice\\
\textbf{travelling} & travel, trip, destin, flight, vacat, book, deal, airlin, hotel, search\\
\bottomrule
\end{tabular}\label{tb:topics} 
\end{table}

\section{Experiments}\label{sec:experiments}

We ran our experiments using the AOL and MSN query logs described in Sec.~\ref{sec:datasets}. 
For the caching simulations, we considered the scenario of storing the query results in the cache (e.g.,~the first SERP). We took as input the stream of queries. If the query is found in the cache we have a cache hit; otherwise, there is a cache miss. In case of miss, if the query is not filtered out by the admission policy implemented (if any) and the dynamic portion of the cache is full, the eviction policy is applied to vacate space for the query results.

For our experiments, we set the cache size $N$ to different values: $64K$, $128K$, $256K$, $512K$, and $1024K$. 
Given the small size of our query logs and the training period needed for learning the topics and the cache initialization, we simulated for our experiments small- and medium-sized caches. In reality, the caches employed by modern search engines are much bigger, but they are always too small compared to the number of queries submitted by the users.

The data in the training set are used for three  purposes: (1) learning the frequency of the queries and loading popular queries in the static cache(s), (2) training the LDA topic classifier from the queries (and clicked documents) and estimating their popularity for balancing the entries of the topic cache(s), and (3) warming up the LRU cache(s). We assess the cache performance in terms of hit rate, namely the number of cache hits in the test set divided by the number of queries in the test set. 

For our experiments, we considered  the following cache organizations:
\begin{itemize}
    \item \sdc: we used as baseline the traditional static and dynamic cache, where the dynamic portion is managed by \lru. This approach is considered the state-of-the-art technique for query-result caching as also confirmed by recent works on caching that used \sdc as a principal baseline (e.g.,~\cite{Kucukyilmaz:IPM2017}).
    \item \stdclruf: the \stdc cache where the topic cache is managed by \lru and all topics receive the same amount of entries.
    \item \stdclruv: the \stdc cache where the topic cache is managed by \lru and the topics receive an amount of entries which is proportional to the topic popularity, as explained in Sec.~\ref{sec:methodology}.
    \item \stdcsdcv: the \stdc cache where the topic cache is managed by \sdc and whose size depends on the popularity of the topic. Compared to the previous two configurations, this cache has another parameter $f_t^s$ representing the static fraction of the \sdc used inside the topic cache. In our tests, we include two different implementations of this configuration. In the first implementation, the static cache $\mathcal{S}$ stores only the frequent queries with no topic (C1) as the popular queries assigned to a topic would be stored in the static portion of the corresponding topic cache. In the second implementation (C2), the $\mathcal{S}$ stores all the top queries (with or without the topic). For popular queries with topic, the algorithm checks if they are already in $\mathcal{S}$. If not, it stores them in the $f_t^s$ fraction of entries of the corresponding \sdc used in topic cache.
    \item \tstdc: The cache entries are divided proportionally to the topic popularity and the no-topic queries that belong to an additional topic $\tau = t_{k+1}$.
\end{itemize}

For the baseline \sdc cache and the proposed \stdc cache configurations, the static parameter $f_s$ varies from 0.0 to 1.0 with step of 0.1, while the other parameters ($f_t$ and $f_d$) are  tuned based on the remaining size of the whole cache, e.g.,~$N \cdot (1- f_{s})$. Regarding \stdcsdcv, the fraction of static  of the \sdc caches used in the topic portion, $f_t^s$, is the same for all the topics. We also experimented with variable $f_t^s$ estimates \emph{per topic}, but the overall experimental results are similar to those achieved with a fixed $f_t^s$, and we do not report them here.

We investigate the following research questions:
\begin{description}
    \item[RQ1.] For a given cache size, is our proposed \stdc cache able to improve the hit rate performance metric w.r.t. \sdc and, if so, adopting which configuration and optimal parameter values?
    \item[RQ2.] Given the best \stdc competitor identified in RQ1, what is the impact of the other configuration parameters? In particular, given a static fraction $f_s$, what is the impact of the topic and dynamic caches of \stdc w.r.t. the dynamic cache of \sdc?
    \item[RQ3.] How large are the hit rate improvements of the best \stdc configuration w.r.t. \sdc, measured in term of the distance with the hit rate of a theoretical optimal caching strategy?
    \item[RQ4.] Do the query admission policies affect the caching performance? How do these policies impact performance gains of \stdc over \sdc?
\end{description}

\begin{table}[bp!]
\centering
\caption{Best hit rates for \sdc and different topic-cache strategies for the AOL and MSN datasets. The parameters $f_s$, $f_t$, and $f_d$ denote the fractions of the total cache devoted to the static, topic, and dynamic caches, respectively. The parameter $f_t^s$ denotes the percentage of static cache in the topic cache. The best hit rates are highlighted in bold.}
\label{tb:hit_ratios}
\begin{adjustbox}{max width=\textwidth}
\begin{tabular}{cccccccccccc} 
\toprule
\multirow{2}{*}{\textbf{Cache Size}} & \multirow{2}{*}{\textbf{Strategy}} & \multicolumn{5}{@{}c@{}}{AOL} & \multicolumn{5}{@{}c@{}}{MSN}\\
\cmidrule(lr){3-7}\cmidrule(lr){8-12}
 &  & \textbf{Hit Rate} & \textbf{$f_{s}$} & \textbf{$f_{t}$}  & \textbf{$f_{d}$} & \textbf{$f_t^s$} & \textbf{Hit Rate} & \textbf{$f_{s}$}& \textbf{$f_{t}$} & \textbf{$f_{d}$} & \textbf{$f_t^s$}\\
\midrule
\multirow{5}{*}{64K}    & \sdc           & 33.70\%           & 0.9 & --   & --   & --   & 45.23\%            & 0.9 & --   & --   & --\\
                        & \stdclruf      & 36.91\%           & 0.9 & 0.07 & 0.03 & --   & 46.57\%            & 0.8 & 0.13 & 0.07 & --\\
                        & \stdclruv      & \textbf{37.34\%}  & 0.9 & 0.07 & 0.03 & --   & 47.09\%           & 0.9 & 0.05 & 0.05 & --\\
                        & \stdcsdcv (C1) & 36.21\%           & 0.8 & 0.16 & 0.04 & 90\% & 46.00\% & 0.1 & 0.72 & 0.18 & 90\%\\
                        & \stdcsdcv (C2) & \textbf{37.34\%}  & 0.8 & 0.16 & 0.04 & 60\% & \textbf{47.15\%}   & 0.8 & 0.13 & 0.07 & 60\%\\
                        & \tstdc         & 33.16\%           & --  & --   & --   & 90\% & 42.30\%            & --  & --   & --   & 80\%\\
\midrule
\multirow{5}{*}{128K}   & \sdc           & 37.58\%           & 0.9 & --   & --   & --   & 48.15 \%           & 0.9 & --   & --   & --   \\
                        & \stdclruf      & 40.89\%           & 0.9 & 0.08 & 0.02 & --   & 49.73\%            & 0.9 & 0.07 & 0.03 & --   \\
                        & \stdclruv      & \textbf{41.19\%}           & 0.9 & 0.07 & 0.03 & --   & 50.04\%            & 0.9 & 0.07 & 0.03 & --   \\
                        & \stdcsdcv (C1) & 40.08\%  & 0.8 & 0.16 & 0.04 & 90\% & 49.13\%   & 0.1 & 0.72 & 0.18 & 90\%\\
                        & \stdcsdcv (C2) & \textbf{41.19\%}  & 0.9 & 0.08 & 0.02 & 20\% & \textbf{50.08\%} & 0.8 & 0.16 & 0.04 & 80\%\\
                        & \tstdc         & 37.49\%           & --  & --   & --   & 90\% & 46.08\%            & --  & --   & --   & 90\%\\
\midrule
\multirow{5}{*}{256K}   & \sdc           & 41.25\%           & 0.9 & --   & --   & --   & 50.77\%            & 0.9 & --   & --   & -- \\
                        & \stdclruf      & 44.57\%           & 0.9 & 0.08 & 0.02 & --   & 52.50\%            & 0.9 & 0.08 & 0.02 & --   \\
                        & \stdclruv      & 
                        \textbf{44.80\%} & 0.9 & 0.08 & 0.02 & --   & \textbf{52.63\%}            & 0.9 & 0.07 & 0.03 & --   \\
                        & \stdcsdcv (C1) & 43.63\%  & 0.7 & 0.24 & 0.06 & 90\% & 51.94\% & 0.1 & 0.72 & 0.18 & 90\% \\
                        & \stdcsdcv (C2) & \textbf{44.80\%}  & 0.9 & 0.08 & 0.02 & 20\% & \textbf{52.63\%} & 0.9 & 0.07 & 0.03 & 30\%\\
                        & \tstdc         & 41.70\%           & --  & --   & --   & 90\% & 49.46\%            & --  & --   & --   & 90\% \\
\midrule
\multirow{5}{*}{512K}   & \sdc           & 44.52\%           & 0.9 & --   & --   & --   & 52.91\%            & 0.9 & -- & -- & --   \\
                        & \stdclruf      & 47.74\%           & 0.9 & 0.07 & 0.03 & --   & 54.60\%            & 0.9 & 0.08 & 0.02& --   \\
                        & \stdclruv      & 48.06\%           & 0.8 & 0.16 & 0.04 & --   & \textbf{54.83\%}            & 0.8 & 0.16 & 0.04 & --   \\
                        & \stdcsdcv (C1) & 46.93\%  & 0.7 & 0.15 & 0.15 & 80\% & 54.43\%   & 0.2 & 0.64 & 0.16 & 90\%   \\
                        & \stdcsdcv (C2) & \textbf{48.08\%}  & 0.8 & 0.16 & 0.04 & 20\% & \textbf{54.83\%}   & 0.8 & 0.16 & 0.04 & 10\%\\
                        & \tstdc         & 45.60\%           & --  & --   & --   & 90\% & 52.52\%            & --  & --   & --   & 90\%   \\
 \midrule
 \multirow{5}{*}{1024K} & \sdc           & 47.37\%           & 0.7 & --   & --   & --   & 54.93\%            & 0.9 & --   & -- & --   \\
                        & \stdclruf      & 50.31\%           & 0.8 & 0.13 & 0.07 & --   & 56.57\%            & 0.9 & 0.07 & 0.03 & --   \\
                        & \stdclruv      & 50.90\%  & 0.6 & 0.32 & 0.08 & --   & 56.86\%            & 0.7 & 0.24  & 0.06 & --   \\
                        & \stdcsdcv (C1) & 49.98\%           & 0.4 & 0.40 & 0.20 & 80\% & 56.72\%   & 0.2 & 0.64 & 0.16 & 80\%   \\
                        & \stdcsdcv (C2) & \textbf{51.01\%} & 0.5 & 0.40 & 0.10 & 70\% & \textbf{56.92\%}   & 0.5 & 0.40 & 0.10 & 50\%\\
                        & \tstdc         & 49.33\%           & --  & --   & --   & 80\% & 55.16\%            & --  & --  & --& 90\%   \\
\bottomrule
\end{tabular}
\end{adjustbox}
\end{table}

To address RQ1, we assumed a cache with a given number of entries (e.g.,~$N$~is defined by the system administrator), and we aim to discover the best cache configuration and parameters in terms of hit rate. 
Table~\ref{tb:hit_ratios} reports the best hit rates obtained with \sdc (our baseline) and the other topic-caching strategies for different cache sizes. For each caching strategy, we report also the values of the $f_s$, $f_t$, $f_d$, and $f_t^s$ that achieved the best hit rates. As shown, not all parameters are used by all the cache configurations, so for those caches where the parameter is not needed we use the symbol~$-$. 
For each cache size, the best hit rates are highlighted in bold. Our experiments showed that with both datasets the \stdc caches always perform better than \sdc in terms of hit rate. In particular, the \stdcsdcv (C2) configuration performs always better than the others.
As expected, \stdclruf performs worse than \stdclruv as it gives to each topic the same number of entries instead of allocating the topic cache entries proportionally to the topic popularity. Moreover, \stdcsdcv (C1) cache exhibits lower hit rates compared to \stdcsdcv (C2) and \stdclruv caches. Analyzing the cache misses encountered with (C1), we  see that this reduction of performance is due to the fact that the static cache $\mathcal{S}$ of (C1) hosts only the results of \textit{no-topic} queries. Some of these queries may be not very popular, hence storing them in the static fraction causes a lower hit rate in static with a reduction of the overall performance. In particular, this phenomenon is more evident when $f_s$ increases, since we are allocating more space to $\mathcal{S}$ and, at some point, also infrequent \textit{no-topic} queries are selected just to fill in the space. Nevertheless, (C2) does not suffer from this, since it stores in $\mathcal{S}$ the frequent queries (with or without topic), allowing a better utilization of the static fraction of the whole cache.  
We could also observe that the \tstdc cache has lower performance than the other \stdc configurations. In most of the cases it performs close to \sdc, and for small caches it does not improve the baseline. Anyway, we believe that its results allow us to better understand the benefit of using a topic cache together with static and dynamic caches. In \stdclruf,  \stdclruv, and \stdcsdcv the amount of entries dedicated to the \textit{no-topic} queries is limited by the parameter $f_d$. Hence, there is a fair division of the cache space among the queries belonging to a topic and those that could not be classified. On the other hand, in \tstdc the \textit{no-topic} queries are treated as queries belonging to an extra topic (i.e., $t_{k+1}$), so the amount of entries is proportional to the popularity of the $(k+1)$-th topic, penalizing the other $k$ topics. Since in our data most of the queries are not classified, this leads to an unbalanced splitting of the space between the \textit{no-topic} queries and the others.

\begin{figure}[t!]
    \centering
    \includegraphics[width=.55\textwidth]{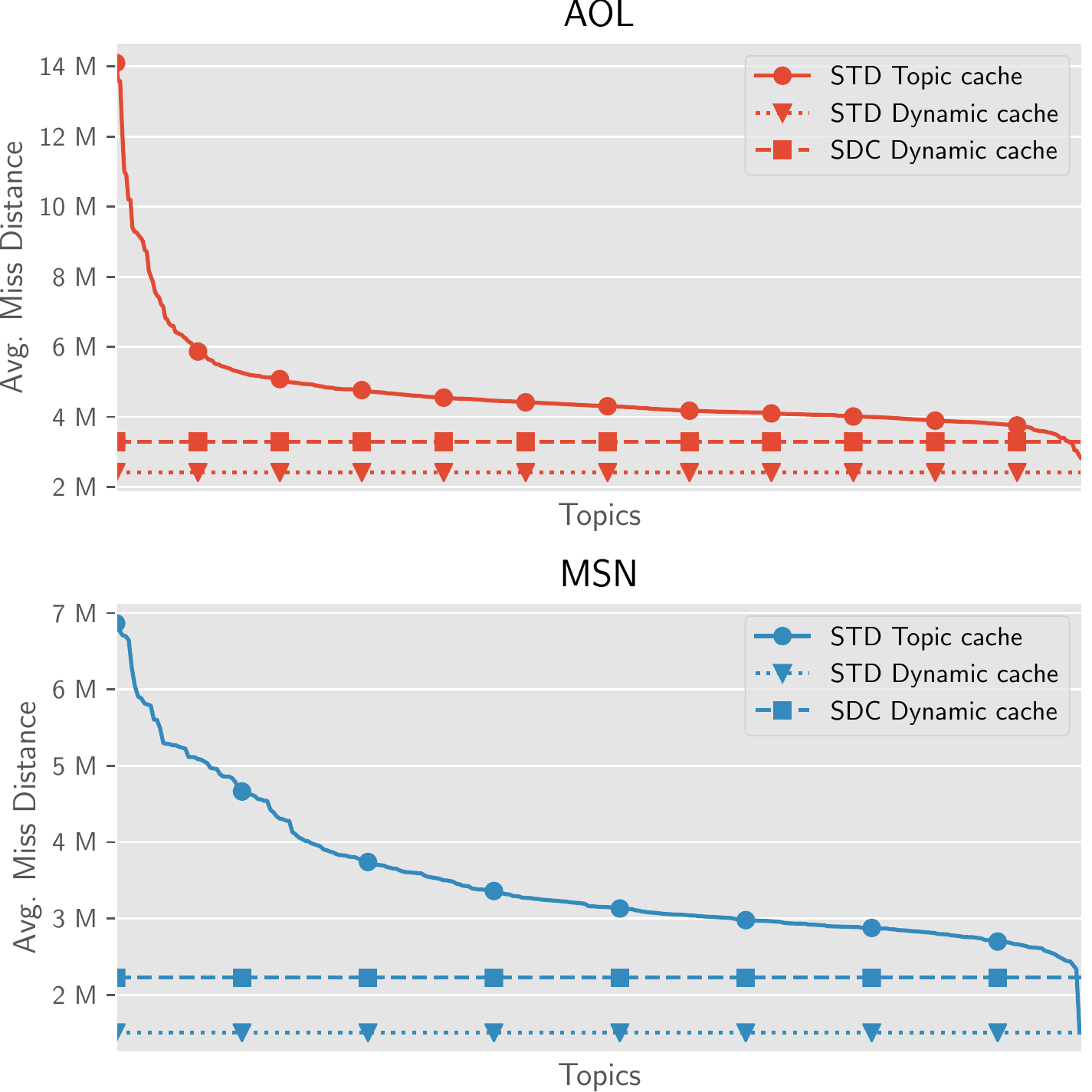}
    \caption{Per-topic average miss distances for AOL and MSN query logs.}
    \label{fig:distances}
\end{figure}

We investigated the reasons why our \stdc cache outperforms the \sdc baseline. Since the higher \stdc hit rate is due to less misses encountered, we analyzed the average distance of misses in the test streams (\textit{avg. miss distance}). This distance is defined as the number of queries between two misses that were caused by the same query, e.g.,~for the stream \texttt{abcadafga} and a cache of size 2 the misses caused by \texttt{a} have an average distance of 2. 
For this experiment, we considered caches with 1024K entries and $f_s = 0.5$, as it gave the best hit rate performance for \stdcsdcv (the best configuration). We separately identified the avg. miss distance of its dynamic cache from the avg. miss distance of topic caches in \stdcsdcv. Notice that the static cache does not impact on the analysis since it is populated by the same top-frequent queries for both \stdc and \sdc caches. 
The results are shown in Fig.~\ref{fig:distances}. The curves represent the avg. miss distances for the topic caches sorted by decreasing values, and we use it as a proxy of temporal locality. On the left we have large distances, which means that a miss occurred only when the repeated requests of that query were far away from each other. Notice that the number of topics can be lower than $500$ since for some topics there are no misses. The avg. miss distances for the dynamic caches in \sdc and \stdc are constant as they are topic-independent. These two avg. miss distances are lower when compared to those reported for most of the topic caches. It confirms that a \lru dynamic cache captures the repeated requests only if they are are close to each other (small avg. miss distance). On the other hand, topic caches have large avg. miss distances. So, the advantage of a topic cache with space divided in a proportional way among the topics is that it even serves requests distant from each other on a per-topic base, i.e.,~with different temporal localities.

To conclude on RQ1, the experimental results confirmed that, on equal cache sizes, the \stdc approach can improve \sdc, resulting in an improvement of the hit rate up to ${\sim}3.6\%$ for AOL and ${\sim}2\%$ for MSN. As confirmed by simulations on two real-world query logs, the best configuration is \stdcsdcv (C2), although \stdclruv performs close.
Moreover, this performance improvement is justified by the analysis of the avg. miss distances. In fact, in \stdc the misses occurring in the topic caches are caused by repeated requests that are much more distant in the query stream as compared to the misses that are encountered in \sdc.

To address RQ2, and see if the improvement of \stdcsdcv (C2) over \sdc is consistent, we compared their hit rates varying the cache size and the value of $f_s$. Since the size of the static portion changes with $f_s$, we split the remaining $N \cdot (1 - f_s)$ entries between the topic and dynamic caches using different proportions. Here, we report the results obtained with $80\%$ for the topic cache and $20\%$ for the dynamic cache, while the $f_t^s$ parameter was set to $40\%$. Consistent results were observed for other parameter values.

\begin{figure*}[tb]
\centering
\includegraphics[width=0.95\textwidth]{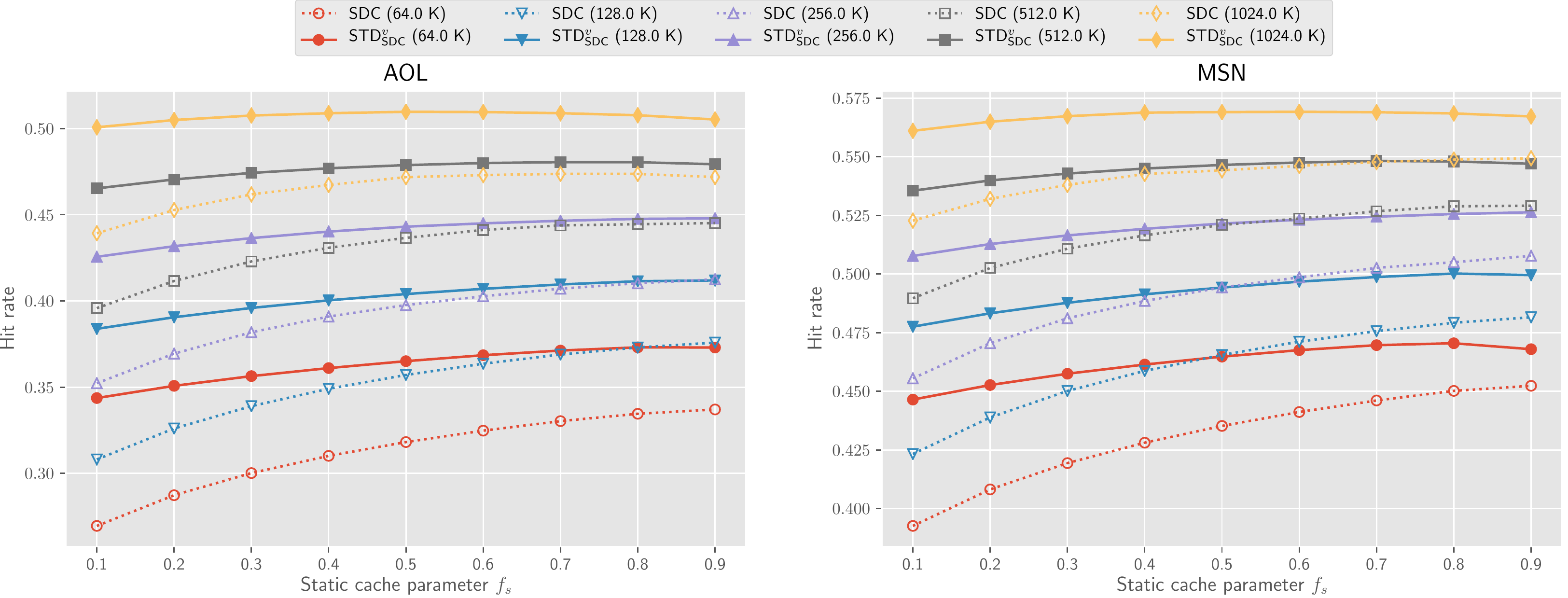}
\caption{Hit rates of \sdc and \stdcsdcv for different values of $N$ and of $f_s$, for the AOL (left) and MSN (right) query logs.}
\label{fig:results}
\end{figure*}

In Fig.~\ref{fig:results}, we illustrate the hit rates for the two approaches, using dashed lines for \sdc and solid lines for \stdcsdcv (C2). Notice that we omit the hit rates for $f_s=0.0$ and $f_s=1.0$ as they correspond to completely dynamic and static caches, and the performance among the approaches are the same. 
If we observe the red curves for $N=64K$, \sdc hit rates (dashed lines) are always lower than \stdcsdcv hit rates (solid line). The gap of hit rates between these two caching approaches goes from ${\sim}7\%$ for $f_s=0.1$ to ${\sim}3\%$ for $f_s=0.9$ for AOL and ${\sim}5\%$ for $f_s=0.1$ to ${\sim}2\%$ for $f_s=0.9$ for MSN. As expected, the maximum improvement is registered for lower values of $f_s$, since the impact of a topic plus a dynamic cache of \stdc over the only dynamic cache of \sdc is more evident. We  observed similar results for the other cache sizes. Regarding RQ2, we can conclude that the \stdcsdcv cache always outperforms the \sdc cache, with an average gap of $5\%$ for AOL and $3\%$ for MSN, and a maximum gap of more than $5\%$ on both query logs.

To answer to RQ3, we compared the best hit rates achieved with \stdc and \sdc against the best hit rate that can be achieved with an optimal cache policy. We used the B\'el\'ady's optimal algorithm (also known as the clairvoyant algorithm), which always evicts the element that will not be requested for the longest time. It is not feasible in practice as it assumes to know the future requests, but it optimizes the number of hits~\cite{Belady:1969}, and it gives us an upper bound of the performance over which no other caching strategy can improve. 
We computed the gaps between B\'el\'ady hit rates and the ones achieved with the best \sdc and \stdc configurations. The results are reported in Table~\ref{tb:improvement}. As we can see, the hit rates of \stdc are always higher than those of \sdc. We report the gap between \stdc and \sdc in the 7th column of the table and the average gap is $3.6\%$ for AOL and $1.9\%$ for MSN (averaged over the size of the cache). Moreover, the hit rates of \stdc are very close to B\'el\'ady hit rates for all cache sizes (the gap is reported in the 6th column of Table~\ref{tb:improvement}). The average gap between the hit rates of \stdc and B\'el\'ady is $6.70\%$ for AOL and $5.06\%$ for MSN (averaged over the size of the cache). On the other hand, the distance between \sdc hit rate and B\'el\'ady hit rate is bigger (see the 5th column). The average gap between them is $10.30\%$ for AOL and $6.98\%$ for MSN. To quantify the gap reduction, we computed the relative delta between the two gaps (see the 8th column). It gives us an indication on how much \stdc improves \sdc w.r.t. B\'el\'ady hit rate. To conclude on RQ3, \stdc hit rates achieve a significant gap reduction w.r.t. \sdc from the theoretical optimal hit rate, which is up to $36.51\%$ for AOL and up to $31.04\%$ for MSN.

\begin{table}[t!]
\centering
\caption{Gap between the best hit rates achieved with \sdc and \stdc caches w.r.t. B\'el\'ady and each other, plus the gap reduction w.r.t. B\'el\'ady for the best hit rates achieved with \sdc and \stdc caches.}
\label{tb:improvement}
\begin{adjustbox}{max width=\textwidth}
\begin{tabular}{cccccccc} 
\toprule
\textbf{Cache Size} & \textbf{B\'el\'ady} & \textbf{Best \sdc} & \textbf{Best \stdc} & \makecell{\textbf{Gap \sdc}\\\textbf{w.r.t. B\'el\'ady}} & \makecell{\textbf{Gap \stdc}\\\textbf{w.r.t. B\'el\'ady}} & \makecell{\textbf{Gap  \stdc} \\ \textbf{w.r.t. \sdc}} & \makecell{\textbf{Gap} \\\textbf{Reduction}}\\
\midrule
\multicolumn{8}{@{}c@{}}{AOL}\\
\midrule
64K   & 43.67\% & 33.70\% & 37.34\% & 9.97\% & 6.33\% & 3.64\% & 36.51\% \\
128K  & 47.68\% & 37.58\% & 41.19\% & 10.10\% & 6.49\% & 3.61\% & 35.74\% \\
256K  & 51.67\% & 41.25\% & 44.80\% & 10.42\% & 6.87\% & 3.55\% & 34.06\% \\
512K  & 54.88\% & 44.52\% & 48.08\% & 10.36\% & 6.80\% & 3.56\% & 34.36\% \\
1024K & 58.06\% & 47.37\% & 51.01\% & 10.69\% & 7.05\% & 3.64\% & 34.05\% \\
\midrule
\multicolumn{8}{@{}c@{}}{MSN}\\
\midrule
64K   & 51.54\% & 45.23\% & 47.15\% & 6.31\% & 4.39\% & 1.92\% & 30.42\% \\
128K  & 54.73\% & 48.15\% & 50.08\% & 6.58\% & 4.65\% & 1.93\% & 29.33\% \\
256K  & 57.98\% & 50.77\% & 52.63\% & 7.21\% & 5.35\% & 1.86\% & 25.80\% \\
512K  & 61.33\% & 52.91\% & 54.83\% & 8.42\% & 6.50\% & 1.92\% & 22.80\% \\
1024K & 61.34\% & 54.93\% & 56.92\% & 6.41\% & 4.42\% & 1.99\% & 31.04\% \\
\bottomrule
\end{tabular}
\end{adjustbox}
\end{table}

Lastly, we addressed RQ4 by implementing two admission policies to be   used in conjunction with the \sdc cache and  the different \stdc configurations. The first admission policy tries to predict queries that are not worthy, namely should not be cached, also known as \emph{polluting queries}~\cite{baeza:SPIRE2007admission}. The second one uses an oracle to determine the singleton queries that are not going to be requested in the future, hence they are not admitted in the cache. This allows to have an upper bound of the performance as no other admission policy can do better than the oracle which is able to see the future and determine no longer requested queries.

\paragraph{Caching Without Polluting Queries} Some queries are not worthy to be cached as they are not requested again or they are requested after a long period, generally after their eviction from the cache. To improve caching algorithms, researchers have tried to predict the singleton/polluting queries which are not admitted in the cache, with the purpose of preserving space for other, more promising, queries. Baeza-Yates et al.~\cite{baeza:SPIRE2007admission} analyzed stateless and stateful features for admitting queries in the cache. The stateless features do not require statistics from previous queries and can be computed on-the-fly based on the query characteristics (e.g., the query length as the number of words or  characters). The assumption is that too long queries are not requested by many users, so very likely they will not appear in the future and are not worth caching. The stateful features are based on historical statistics computed over the query log, such as the number of times the query has been already requested.  
Other more sophisticated admission policies are based on machine learning (e.g., regression models) that predict the next query request~\cite{Kucukyilmaz:IPM2017}. They rely on several features of queries (e.g., query length, presence of typos), index (e.g., length of posting lists of most common or uncommon terms), query and term frequencies as well as user's sessions (e.g., user logged in, clickthrough rate). The intent is to admit to the static cache only those queries with high expected frequency and to the dynamic cache the queries with high probability to be resubmitted. The authors tried different cache configurations and showed an improvement of $0.66\%$ for static cache and of $0.47\%$ for \sdc when their admission policy is used. We did not implement this admission policy as we lack information for index and session features, but our improvement was higher even with the admission policy based on query features (e.g.,~\cite{baeza:SPIRE2007admission}). Besides, in the next paragraph, we show hit rates of \stdc and \sdc using an oracle that prevents singleton queries from entering the cache, and this can be seen as a performance upper bound for admission techniques.

Following~\cite{baeza:SPIRE2007admission}, we implemented an  admission strategy that accepts a query in the cache based on stateless and stateful features. In particular, the query is cached only if it satisfies the following conditions:

\begin{itemize}
    \item it has been requested in the training period at least X times (stateful feature);
    \item the number of terms in the query is less than Y (stateless features);
    \item the number of characters is less than Z (stateless features);
\end{itemize}

For the experiments discussed in this paper, we set $X = 3$, $Y = 5$, and $Z = 20$. 
Similar results were achieved with other threshold values proposed in~\cite{baeza:SPIRE2007admission}. For these experiments with the admission policy, we decide to use the split $30\%-70\%$ of data in order to have more queries in the test set for the evaluation. Indeed, we observed that in the absence of polluting queries, using only $30\%$ for test set results in a number of test queries which is too small for assessing the larger caches (e.g., 512K or 1024K entries). 
The results of these experiments are reported in Table~\ref{tb:hit_ratios_without_polluting}. As we can see, the admission policy improves the hit rates of both \sdc and \stdc caches. Note that for all configurations of cache sizes and datasets, \stdc outperforms \sdc. 

\begin{table}[t!]
\centering
\caption{Hit rates achieved with \sdc and \stdc caches with an admission policy that does not admit the polluting queries into the cache. The best hit rates are highlighted in bold.} 
\label{tb:hit_ratios_without_polluting}
\begin{tabular}{ccccccc} 
\toprule
\textbf{Cache Size} & \textbf{\sdc} &  \textbf{\stdclruf} & \textbf{\stdclruv} & \textbf{\stdcsdcv (C1)} & \textbf{\stdcsdcv (C2)} & \textbf{\tstdc} \\
\midrule
\multicolumn{7}{@{}c@{}}{AOL}\\
\midrule
64K & 37.98\% & 41.53\% & 41.85\% & 40.07\% & \textbf{41.87\%} & 37.05\% \\
128K & 40.40\% & 44.08\% & 44.38\% & 43.28\% & \textbf{44.45\%} & 41.21\% \\
256K & 41.48\% & 44.86\% & \textbf{45.52\%} & 44.96\% & 45.51\% & 43.84\% \\
512K & 42.58\% & 45.74\% & \textbf{46.72\%} & 46.39\% & 46.59\% & 45.74\%\\
1024K & 43.91\% & 46.69\% & \textbf{48.02\%} & 47.82\% & 47.88\% & 47.43\%\\
\midrule
\multicolumn{7}{@{}c@{}}{MSN}\\
\midrule
64K   & 48.35\% & 49.85\% & 50.27\% & 49.06\% & \textbf{50.29\%} & 45.47\%\\
128K & 48.83\% &50.82\%  & \textbf{51.01\%} & 50.57\% & 50.97\% & 48.70\%\\
256K & 49.22\% & 51.16\% & \textbf{51.63\%} & 51.40\% & 51.61\% & 50.45\%\\
512K & 49.86\% & 51.58\% & \textbf{52.45\%} & 52.33\% & 52.39\% & 51.56\%\\
1024K & 50.84\% & 52.21\% & \textbf{53.27\%} & 53.19\% & 53.20\% & 52.84\%\\
\bottomrule
\end{tabular}
\end{table}

\begin{table}[t!]
\centering
\caption{Caching with an admission policy that removes the polluting queries: gap between the best hit rates achieved with \sdc and \stdc caches w.r.t. B\'el\'ady and each other, plus the gap reduction w.r.t. B\'el\'ady hit rate for the best hit rates achieved with \sdc and \stdc caches.}

\label{tb:improvement_without_pollution}
\begin{adjustbox}{max width=\textwidth}
\begin{tabular}{cccccccc} 
\toprule
\textbf{Cache Size} & \textbf{B\'el\'ady} & \textbf{Best \sdc} & \textbf{Best \stdc} & \makecell{\textbf{Gap \sdc}\\\textbf{w.r.t. B\'el\'ady}} & \makecell{\textbf{Gap \stdc} \\ \textbf{w.r.t. B\'el\'ady}} & \makecell{\textbf{Gap \stdc} \\ \textbf{w.r.t. \sdc}} & \makecell{\textbf{Gap}\\ \textbf{Reduction}}\\
\midrule
\multicolumn{8}{@{}c@{}}{AOL}\\
\midrule
64K & 46.37\% & 37.98\% & 41.87\% & 8.39\% & 4.50\% & 3.89\% & 46.36\%\\
128K & 49.13\% & 40.40\% & 44.45\% & 8.73\% & 4.68\% & 4.05\% & 46.39\% \\
256K & 50.90\% & 41.48\% & 45.52\% & 9.42\% & 5.38\% & 4.04\% & 42.88\% \\
512K & 51.29\% & 42.58\% & 46.72\% & 8.71\% & 4.57\% & 4.14\% & 47.53\% \\
1024K & 51.30\% & 43.91\% & 48.02\% & 7.39\% & 3.28\% & 4.11\% & 55.61\% \\
\midrule
\multicolumn{8}{@{}c@{}}{MSN}\\
\midrule
64K   & 52.81\% & 48.35\% & 50.29\% & 4.46\% & 2.52\% & 1.94\% & 43.49\%\\
128K & 54.40\% & 48.83\% & 51.01\% & 5.57\% & 3.39\% & 2.18\% & 39.13\%\\
256K & 55.11\% & 49.22\% & 51.63\% & 5.89\% & 3.48\% & 2.41\% & 40.91\%\\
512K & 55.11\% & 49.86\% & 52.45\% & 5.25\% & 2.66\% & 2.59\% & 49.33\%\\
1024K & 55.11\% & 50.84\% & 53.27\% & 4.27\% & 1.84\% & 2.43\% & 56.90\%\\
\bottomrule
\end{tabular}
\end{adjustbox}
\end{table}

Table~\ref{tb:improvement_without_pollution} reports the gaps between hit rates achieved by \sdc and by the best \stdc configuration with respect to the hit rates of B\'el\'ady algorithm. In all these experiments, we restrict polluting queries from entering the cache. We also computed the gap between \sdc and \stdc (see 7th column of the table) and the gap improvement w.r.t. B\'el\'ady (see 8th column of the table). 
We observe that \stdc continues to improve \sdc with larger gaps compared to the caches without admission policy (see the 7th column of Table~\ref{tb:improvement}), reaching peaks of $4.14\%$ for  AOL and of $2.59\%$ for MSN. The gap improvement w.r.t. B\'el\'ady is more than $55\%$ for both datasets.

\begin{figure}[t!]
    \centering
    \includegraphics[width=.95\textwidth]{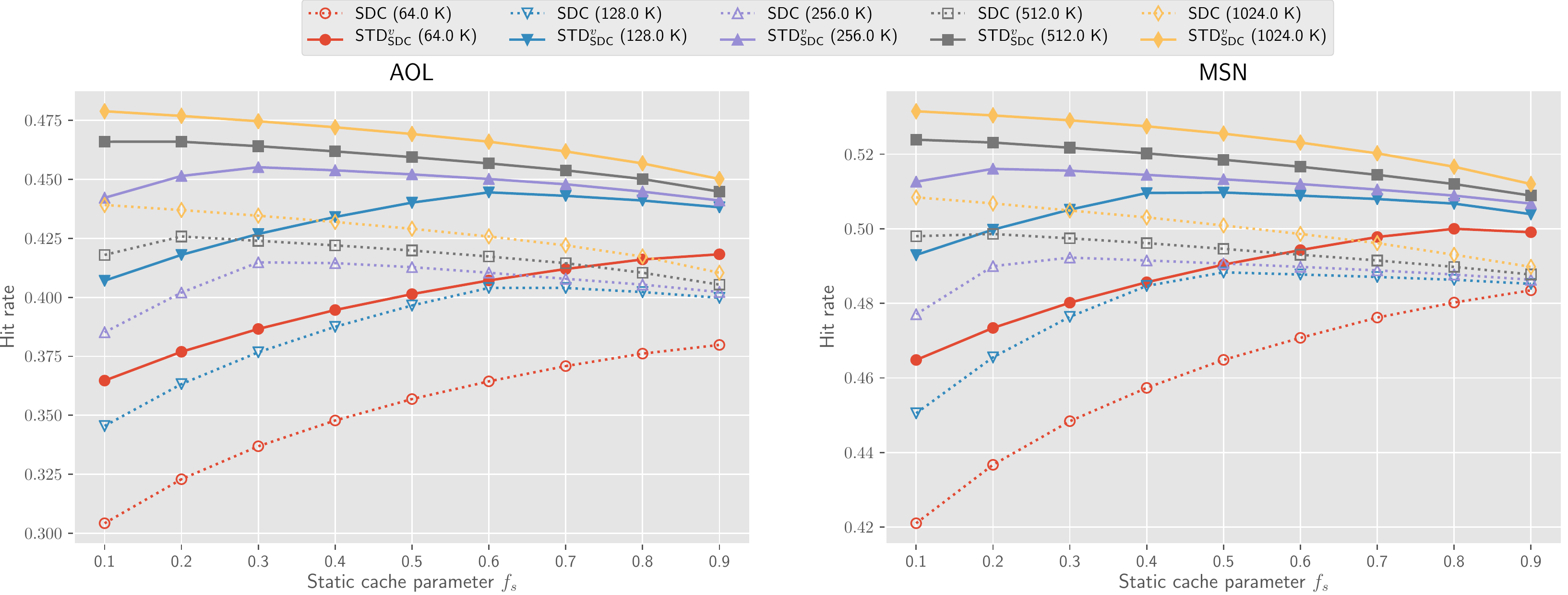}
    \caption{Hit rates of \sdc and \stdcsdcv, using the admission policy that removes polluting queries, for different values of $N$ and of $f_s$, for the AOL (left) and MSN (right) query logs.}
    \label{fig:hit_rates_without_polluting}
\end{figure}

In Fig.~\ref{fig:hit_rates_without_polluting}, we show the hit rates for \sdc and \stdc with the admission policy that does not allow caching of polluting queries. We remind the reader that the dashed lines are for \sdc and solid lines for \stdcsdcv (C2), and we show results for the static fraction $f_s$ varying from 0.1 to 0.9. Each pair of curves represent a different size of the cache. As we can see, for the same cache size and $f_s$ value, the \stdcsdcv (C2) hit rate curve is always higher than the \sdc curve.

\paragraph{Caching Without Singleton Queries} Since \stdc uses topical information to better cache query results, one could surmise that the \stdc cache capitalizes on non-singleton queries as they were observed in the training period and consequently have a topic assigned. To  show that the advantage encountered with \stdc is irrespective of non-singleton queries, we experimented with an admission policy that restricts singleton queries from entering the cache.

A singleton query, by definition, is requested only once; hence, caching its results wastes space as the query will not be requested again. 
We implemented an oracle (i.e., an algorithm that knows the future) which, given as input the query stream, returns the singleton queries (i.e., queries that appear only once in the stream). This is obviously not feasible in practice but does provide an optimal environment to compare \stdc and \sdc. Then, we ran our experiments not admitting in the caches these singleton queries. 

As we did for the experiment on polluting queries, also for this experiment we decided to use the split $30\%-70\%$ to have more queries in the test set. We show the results in  Table~\ref{tb:hit_ratios_without_singleton}. Again, the \stdc approach always improves \sdc in terms of hit rates. Overall, the hit rates obtained without singleton queries are much higher as compared to those of Table~\ref{tb:hit_ratios} since not caching singleton queries spares space for future worthy queries.

\begin{table}[t!]
\centering
\caption{Hit rates achieved with \sdc and \stdc caches with a perfect admission policy that does not allow the singleton queries. The best hit rates are highlighted in bold.} 
\label{tb:hit_ratios_without_singleton}
\begin{tabular}{ccccccc} 
\toprule
\textbf{Cache Size} & \textbf{\sdc} &  \textbf{\stdclruf} & \textbf{\stdclruv} & \textbf{\stdcsdcv (C1)} & \textbf{\stdcsdcv (C2)} & \textbf{\tstdc} \\
\midrule
\multicolumn{7}{@{}c@{}}{AOL}\\
\midrule
64K & 57.27\% & 59.88\% & 60.47\% & 58.47\% & \textbf{60.48\%} & 54.31\% \\	
128K &	63.47\% & 65.72\%  & 66.38 \% & 65.01\% & \textbf{66.39\%} & 61.05\% \\	
256K &	70.28\% & 71.27\%&	72.81\% & 71.92\% & \textbf{72.83\%} & 68.03\% \\
512K &	77.80\% & 76.32\% & \textbf{79.87\%} & 79.45\% & 79.84\% & 75.84\% \\ 	
1024K &	86.66\% &	 82.27\% & \textbf{87.73\%}  & 87.62\% & 87.69\% & 84.49\%  \\	
\midrule
\multicolumn{7}{@{}c@{}}{MSN}\\
\midrule
64K & 67.85\% & 68.85\% & 69.41\% & 68.06\% & \textbf{69.44\%} & 63.54\% \\	
128K & 71.64\% & 72.83\% & \textbf{73.24\%} & 72.37\% & 73.22\% & 68.49\%\\	
256K & 75.60\% & 76.16\% & \textbf{77.41\%} & 76.80\% & 77.35\% & 73.14\%\\	
512K & 80.89\% & 79.54\% & \textbf{82.80\%} & 82.55\% & 82.72\% &  78.72\%\\	
1024K & 87.18\% & 83.70\% & \textbf{87.90\%} & 87.75\% & 87.85\% & 85.33\%\\	
\bottomrule
\end{tabular}
\end{table}

In Table~\ref{tb:improvement_without_singleton} we report the gaps between the hit rates of \sdc and best \stdc caches with respect to B\'el\'ady as well as the gap between \sdc and best \stdc (see the 7th column of the table). We observe that the gaps are smaller compared to the gaps of caches without admission policies (see the 7th column of Table~\ref{tb:improvement}), especially for large caches. This is expected, as both caches are advantaged by removing singleton queries. The gap reduction is also smaller than the one observed in the other experiments as \sdc and \stdc  perform close to the optimum. Nevertheless, this experiment proved that \stdc  still has higher hit rates as compared to \sdc. We remind the reader that this admission policy is unfeasible in practice since the oracle  assumes to know the future. Anyway, we included these results to show how \stdc improves \sdc even in extreme cases where the caching approaches have the advantage of not storing singleton queries. Moreover, \stdc does not benefit by not storing the non-singleton queries more than \sdc could do. Finally, this result is a good indication that \sdc even with the best admission strategy (i.e., the oracle in our case) on the top of it cannot perform better than \stdc.

\begin{table}[t!]
\centering
\caption{Caching with an admission policy that removes the singleton queries: Gap between the best hit rates achieved with \sdc and \stdc caches, plus the gap reduction w.r.t. B\'el\'ady hit rate for the best hit rates achieved with the admission policy.}
\label{tb:improvement_without_singleton}
\begin{adjustbox}{max width=\textwidth}
\begin{tabular}{cccccccc} 
\toprule
\textbf{Cache Size} & \textbf{B\'el\'ady} & \textbf{Best \sdc} & \textbf{Best \stdc} & \makecell{\textbf{Gap \sdc}\\\textbf{w.r.t. B\'el\'ady}} & \makecell{\textbf{Gap Best \stdc} \\ \textbf{w.r.t. B\'el\'ady}} & \makecell{\textbf{Gap \sdc} \\ \textbf{and Best \stdc}} & \makecell{\textbf{Gap}\\ \textbf{Reduction}}\\
\midrule
\multicolumn{8}{@{}c@{}}{AOL}\\
\midrule
64K & 68.76\% & 57.27\% & 60.48\% & 11.49\% & 8.28\% & 3.21\% & 27.93\%\\
128K & 75.04\% & 63.47\% & 66.39\% & 11.57\% & 8.65\% & 2.92\% & 25.23\% \\
256K & 81.09\% & 70.28\% & 72.83\% & 10.81\% & 8.26\% & 2.55\% & 23.59\% \\
512K & 85.26\% & 77.80\% & 79.87\% & 7.46\% & 5.39\% & 2.07\% & 27.74\% \\
1024K & 87.88\% & 86.66\% & 87.73\% & 1.22\% & 0.15\% & 1.07\% & 87.70\% \\
\midrule
\multicolumn{8}{@{}c@{}}{MSN}\\
\midrule
64K   & 75.10\% & 67.85\% & 69.44\% & 7.25\% & 5.66\% & 1.59\% & 21.93\%\\
128K & 79.69\% & 71.64\% & 73.24\% & 8.05\% & 6.45\% & 1.60\% & 19.87\%\\
256K & 83.76\% & 75.60\% & 77.41\% & 8.16\% & 6.35\% & 1.81\% & 22.18\%\\
512K & 85.21\% & 80.89\% & 82.80\% & 4.32\% & 2.41\% & 1.91\% & 44.21\%\\
1024K & 88.21\% & 87.18\% & 87.90\% & 1.03\% & 0.31\% & 0.72\% & 69.90\%\\
\bottomrule
\end{tabular}
\end{adjustbox}
\end{table}

\begin{figure}[t!]
    \centering
    \includegraphics[width=.95\textwidth]{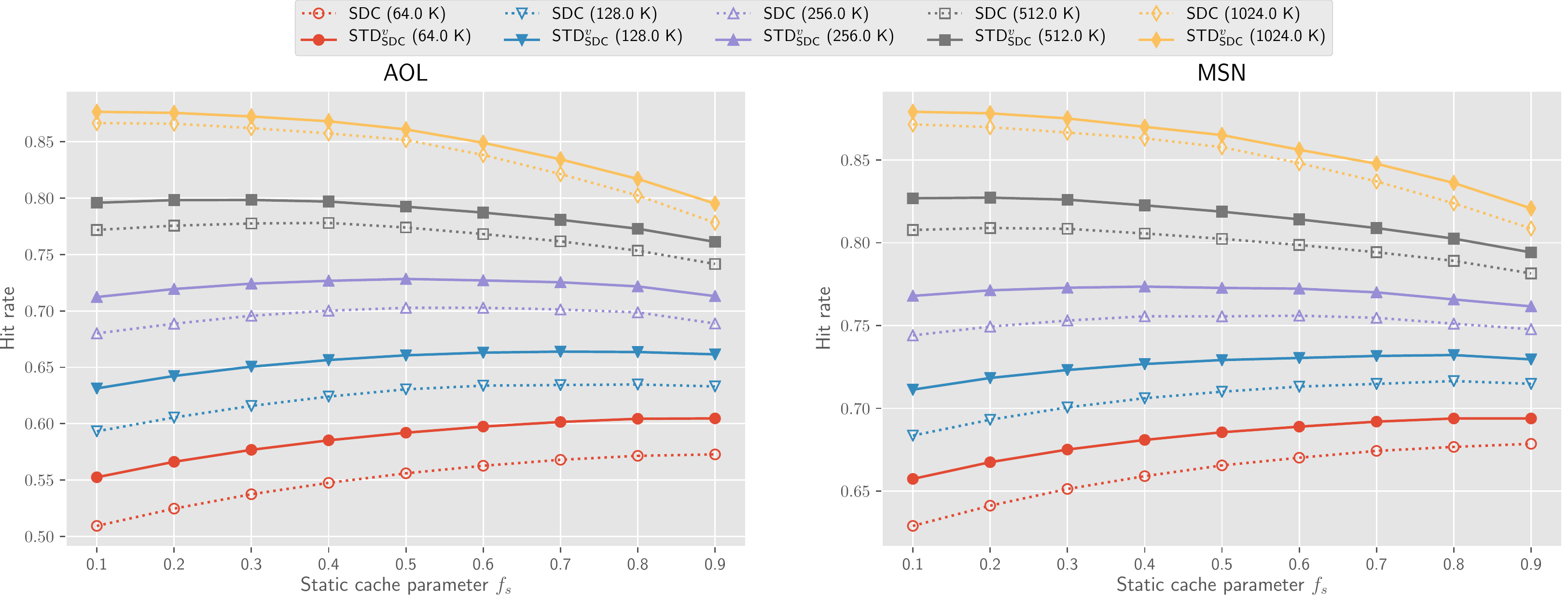}
    \caption{Hit rates of \sdc and \stdcsdcv, using the admission policy that does not allow singleton queries, for different values of $N$ and of $f_s$, for the AOL (left) and MSN (right) query logs.}
    \label{fig:hit_rates_without_singleton}
\end{figure}

In Fig.~\ref{fig:hit_rates_without_singleton} we show the hit rates for \sdc and \stdc with the admission policy that does not allow caching of singleton queries. Also, for this comparison we can see that the performance of \stdcsdcv (C2) are always better than \sdc, although the gap is lower because both remove all singleton requests getting a huge advantage for both caching strategies.

\section{Conclusions}
\label{sec:conclusions}

We presented a novel cache model, Static-Topic-Dynamic (\stdc) cache, which leverages  query topics to better utilize cache space, yielding improved hit rates. Compared to the traditional \sdc cache, the \stdc cache stores queries belonging to a given topic in a dedicated portion of the cache where for each topic the number of entries available is proportional to the topic popularity. This allows to capture queries that are frequently requested at large intervals of time and would be evicted in a cache only managed by the \lru policy. Extensive experiments conducted with two real-world query logs show that  \stdc increases the cache hit rate by more than $3\%$ over \sdc. Such improvements result in a hit rate  gap reduction w.r.t.   B\'el\'ady's optimal caching policy~\cite{Belady:1969} of up to ${\sim}36\%$ over \sdc, depending on the query log and the total size of the cache. The improvement is even higher when an admission policy for not storing polluting queries is employed.

It is worth noting that the greater hit rate achieved by the proposed query-result cache does not require specific investments by the search engine companies. The query topic classification service is in general already deployed for other purposes~\cite{Ophir05}, while our caching solution is managed entirely by software and can be easily implemented and deployed in existing Web search systems.

In the future, we would like to investigate if and how query topic classification performance  impacts the cache hit ratio. We assigned topics to queries by means of a LDA classification technique  trained on a portion of the query logs. We believe that improving  coverage and accuracy of query classification might be beneficial for the \stdc cache hit ratio. As future work, we plan to employ other topic-modeling techniques which are tailored for short text~\cite{Rashid:IPM2019}.
We would also like to use this cache in synergy with other caches, e.g., those storing posting lists of frequently requested~ terms. 

\section*{References}

\bibliography{biblio}

\end{document}